\documentclass[a4paper,11pt,]{article}
\pdfoutput=1
             
\usepackage{jcappub} 
\usepackage{amsmath}
\usepackage{hyperref}
\usepackage{graphicx}
\usepackage[export]{adjustbox}
\usepackage{booktabs}

\begin{document}

\title{Cosmic ray observations using coincident scintillator and radio signals with the Beamforming Elevated Array for COsmic Neutrinos (BEACON)}

\affiliation[a]{Department of Physics,
Kavli Institute for Cosmological Physics,
University of Chicago, Chicago, IL 60637, USA}

\affiliation[b]{Department of Physics,
Colorado School of Mines,
Golden, CO 80401, USA}

\affiliation[c]{Department of Physics,
Pennsylvania State University,
University Park, PA 16802, USA}

\affiliation[d]{Institute for Gravitation
and the Cosmos, Pennsylvania State University,
University Park, PA 16802, USA}

\affiliation[e]{Sorbonne Universit\'e, CNRS, UMR 7095, Institut d'Astrophysique de Paris,
98 bis bd Arago, 75014 Paris, France}

\affiliation[f]{Instituto Galego de F\'\i sica de Altas Enerx\'\i as IGFAE,
Universidade de Santiago de Compostela,
15782 Santiago de Compostela, Spain}

\affiliation[g]{Faculty of Physics, University of Warsaw,
02-093 Warsaw, Poland}

\affiliation[h]{Department of Astronomy and Astrophysics,
Pennsylvania State University,
University Park, PA 16802, USA}

\affiliation[i]{Department of Astronomy and Astrophysics,
Kavli Institute for Cosmological Physics,
University of Chicago, Chicago, IL 60637, USA}

\affiliation[j]{Department of Physics, The Ohio State University,
Columbus, OH 43210, USA}

\affiliation[k]{Department of Astrophysics/IMAPP,
Radboud University Nijmegen,
6500 GL Nijmegen, The Netherlands}

\affiliation[l]{NIKHEF, Science Park Amsterdam,
1098 XG Amsterdam, The Netherlands}

\affiliation[m]{Department of Physics and Astronomy,
University of Kansas, Lawrence, KS 66045, USA}

\affiliation[n]{Jet Propulsion Laboratory,
California Institute of Technology,
Pasadena, CA 91109, USA}

\affiliation[o]{Enrico Fermi Institute,
University of Chicago, Chicago, IL 60637, USA}

\author[a,1]{Z.~Martin\note{Corresponding author.},}
\emailAdd{zacharymartin@uchicago.edu}

\author[b,c]{J.~Hinkel,}
\author[c,d]{K.~Perna,}
\author[c,d,e]{A.~Zeolla,}
\author[f]{J.~Alvarez-Mu\~{n}iz,}
\author[f]{S.~Cabana-Freire,}
\author[g]{W.~Carvalho~Jr.,}
\author[c,d,h]{A.~Cummings,}
\author[i]{C.~Deaconu,}
\author[j]{K.~Hughes,}
\author[c,d]{R.~Krebs,}
\author[j]{Y.~Liu,}
\author[a]{M.~Michulka,}
\author[k,l]{K.~Mulrey,}
\author[j]{A.~Nozdrina,}
\author[i]{E.~Oberla,}
\author[m]{S.~Prohira,}
\author[n]{A.~Romero-Wolf,}
\author[a,i,o]{A.~G.~Vieregg,}
\author[c,d,h]{S.~A.~Wissel,}
\author[f]{E.~Zas}

\collaboration{\includegraphics[height=17mm]{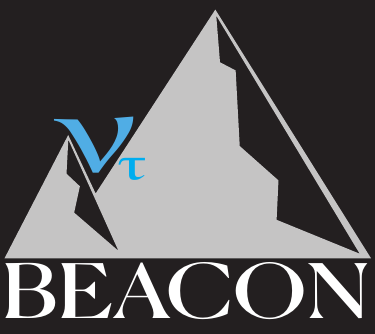}\\[6pt] BEACON Collaboration}


\abstract{
At energies above 100 PeV, coherent radio frequency (RF) emission across the 10 -- 1000 MHz band from cosmic particle-induced atmospheric showers becomes increasingly dominant, making radio detection a powerful probe at ultrahigh energies. The Beamforming Elevated Array for COsmic Neutrinos (BEACON) considers mountaintop phased radio arrays to monitor large areas at the horizon for upward-going geomagnetic emission in the 30 -- 80 MHz band produced as a consequence of Earth-skimming tau neutrinos. As part of ongoing efforts to measure the sensitivity of BEACON to such signals, the BEACON prototype identifies cosmic rays through the same emission mechanism. Prior to 2023, BEACON operated a four-antenna prototype, which successfully demonstrated identification of coherent and impulsive sources using a beamforming RF trigger. In 2023, we upgraded the instrument to six newly designed antennas and added an independent scintillator array to cross-validate radio-detectable cosmic rays. This work details the observation of 22 candidate cosmic-ray events in a scintillator-coincidence-triggered dataset using selection cuts based on expected radio characteristics. A simulation of the hybrid-instrument performance verifies this observation rate to within an order of magnitude. We estimate a background of 0.11 events by applying the selection cuts to synthetic datasets of uncorrelated radio-scintillator pairs that are randomly formed using fixed-rate software-triggered data. The reconstructed signal polarization angles are consistent with pure geomagnetic emission compared to various background classes.
}


\maketitle
\flushbottom


\section{\label{sec:intro}Introduction}

\begin{figure*}[!t]
    \centering
    \includegraphics[width=0.95\linewidth]{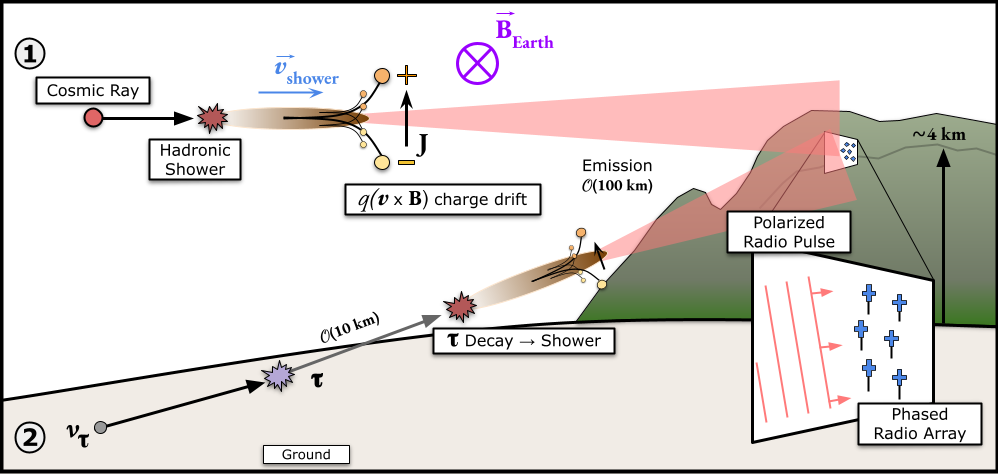}
    \caption{\label{fig:beacondiagram}An overview diagram of the BEACON concept, showcasing emission from extensive air showers initiated by cosmic rays (scenario 1) or Earth-skimming tau neutrinos (scenario 2). The detector consists of a densely packed phased radio array on mountaintops $\sim$4\,km high. The radio mechanism highlighted here is geomagnetic emission produced by a developing transverse current due to the drift of charges along the shower front in the magnetic field of the Earth, although a typically subdominant emission mechanism from a developing frontal charge excess in the atmosphere also contributes to the radio pulse. Note that sensitivity to the tau-neutrino-initiated signal would require $\mathcal{O}$(100 -- 1000) phased array stations.}
\end{figure*}

Observations of cosmic rays (CR) at energies above 10$^{18}$ eV (1 EeV) \cite{EeVCR-Bird_1995, EeVCR-PhysRevD.102.062005} demonstrate the existence of ultrahigh energy (UHE) astrophysical particle accelerators, and imply an associated UHE cosmogenic neutrino flux from UHE CR interactions with cosmic microwave background photons \cite{gzk-PhysRevLett.16.748, gzk-Zatsepin:1966jv}. In addition to the cosmogenic component, direct production of UHE astrophysical neutrinos is predicted in several transient source classes, notably gamma-ray bursts and compact binary mergers \cite{beaconTransModels-Guepin2025Correction, modelsGRB-Murase_2007, modelsBNS-Fang_2017}. Although astrophysical sources are predicted to produce neutrinos predominantly in the electron and muon flavors, a nonzero flux of tau neutrinos is expected due to oscillations from flavor mixing. For standard mechanisms of astrophysical neutrino production, the predicted flavor ratio observed at Earth is approximately even ($\nu_e:\nu_\mu:\nu_\tau = 1:1:1$) \cite{111flavor-PhysRevD.68.093005}; however, certain cases beyond the standard production predict deviations from this (but do not eliminate the tau neutrino flux) \cite{neutrinoFlavor-abbasi2025characterizationthreeflavorcompositioncosmic}.

When such astrophysical particles reach Earth, their interactions in the atmosphere initiate extensive air showers (EAS) that produce coherent impulsive radio emission through geomagnetic charge separation and charge-excess (also called Askaryan) mechanisms, motivating radio experiments for the detection of neutrinos and CRs \cite{emission-Huege_2017, emission-Schr_der_2017, emissionAskaryan-osti_4833087}. Radio measurements additionally provide a nearly continuous duty cycle and sensitivity to the electric field (E-field) footprint of CR EAS emission, enabling reconstruction of the primary energy and the depth of shower maximum, $X_{\text{max}}$ \cite{radioGood-HUEGE20161, radioGood-Buitink_2014, PhysRevLett.132.021001}. More broadly, radio antennas provide a low-cost and scalable detection approach with broad instantaneous sky coverage. At Earth-skimming angles, a UHE tau neutrino can undergo a charged-current interaction in rock near the Earth’s surface, producing a tau lepton that may emerge into the atmosphere, decay, and initiate an upward-going EAS \cite{uptauemission-fargion1999horizontaltauairshowers, uptauemission-PhysRevLett.88.161102, tauGeomagnetic-Zas_2005, emission-Schr_der_2017}.

Several experiments target EAS emission from Earth-skimming tau neutrinos using a range of detector platforms, including ground-based particle and radio detectors \cite{auger-BERTOU2002183, ovrolwa-MONROE2020163086}, mountain and valley geometries \cite{grand-alvarez_Muniz_2019, trinity-Otte_2019, taroge-Wang_2022, magic-AHNEN201877, tambo-tambocollaboration2025tambodeepvalleyneutrinoobservatory, andean-romerowolf2020andeandeepvalleydetectorhighenergy}, balloon-borne instruments \cite{anita-PhysRevD.105.042001, pueo-Abarr_2021}, and satellite concepts \cite{beaconTransPOEMMA-PhysRevD.102.123013, pbr-eser2025poemmaballoonradiooverview}. See ref. \cite{review-Abraham_2022} for a recent review of detection techniques and experiments. Real-time phased-array beamforming with radio-frequency (RF) antennas lowers trigger thresholds for impulsive signals, thereby improving sensitivity to the UHE neutrino flux \cite{nuphasedev, ara-beamform-ALLISON2019112, beamform-PhysRevD.105.122006}.

The Beamforming Elevated Array for COsmic Neutrinos (BEACON) is a proposed mountaintop phased radio detector designed to observe 30 -- 80 MHz radio emission from upward-going EASs initiated by Earth-skimming tau neutrinos with energies above 100 PeV \cite{beaconWhitePaper-Wissel_2020, beacon2021-Southall_2023}. An overview diagram of the instrument with an illustration of the emission mechanisms discussed previously is shown in figure \ref{fig:beacondiagram}. With a clear view of the horizon, high elevation detectors target upward-going EAS emission with an acceptance that increases with detector height \cite{beaconWhitePaper-Wissel_2020}. At full-scale, BEACON will extend this approach to $\mathcal{O}$(100 -- 1000) phased array stations, each comprising $\mathcal{O}$(10) low-cost RF antennas. For stations near the equator and facing towards the east, BEACON has a wide instantaneous field of view that scans 70\% of the sky throughout the duration of a day \cite{beaconSensitivity-Zeolla_2025}. Additionally, an RF beamforming trigger improves signal-to-noise ratio (SNR) by the square root of the number of beamforming antennas, thereby significantly lowering trigger thresholds for faint signals, and narrows triggering to targeted locations, useful for rejecting many RF interference (RFI) backgrounds \cite{beaconWhitePaper-Wissel_2020}. For these reasons, a full-scale BEACON has excellent sensitivity to short-duration point sources \cite{beaconSensitivity-Zeolla_2025}, comparable to that of other next-generation experiments including GRAND, a large-scale ground-based radio array \cite{grand-alvarez_Muniz_2019}, and POEMMA, a space-based observatory \cite{beaconTransPOEMMA-PhysRevD.102.123013}. 

From 2018 to 2023, a BEACON prototype was developed, consisting of four dual-polarized crossed-dipole antennas with an RF trigger implementing real-time beamforming. The prototype spans $\sim 50$ m across the White Mountains of California, near White Mountain Research Center’s (WMRC) Barcroft Field Station. It stands at an altitude of 3.8 km, making it the highest ground-based radio detector for cosmic particles. The array faces eastward, overlooking Fish Lake Valley at an average $\sim 1.6$ km floor and an eastern viewing horizon of $\sim 170$ km, cut short by the eastern mountain ranges (see ref. \cite{beaconSensitivity-Zeolla_2025} for topographical effects on BEACON's sensitivity). 

The prototype informs the performance and sensitivity of BEACON to the tau neutrino radio signature by analyzing the sensitivity to CRs which give a similar radio signature; thus, the analysis of prototype data consists of a CR search in radio. An analysis of data collected by the prototype in 2021 successfully demonstrated the capability of beamforming to trigger on impulsive events, and resulted in a few candidate CR events, although without a robust background estimate \cite{beacon2021-Southall_2023, danthesis-2022PhDT19S}. The beamforming trigger recorded $\sim 1$ million events per day, but simulations predicted $\sim 1$ CR event per day \cite{andrewthesis-Zeolla:2026pan}, thereby requiring significant cuts on the background. A later analysis improved the background estimate using different analysis parameters and a sideband method \cite{abcd-Buttinger2018BackgroundEW}, resulting in one candidate CR event with an estimated background of 0.3 $\pm$ 0.1 events \cite{andrewthesis-Zeolla:2026pan}.

The previous analyses showed that background RFI from anthropogenic noise in towns or from airplanes at this site made it challenging to obtain statistically significant detections of CR events. As a result, we upgraded the instrument in 2023, increasing the number of antennas from four to six and installing an independent scintillator array that could cross-validate the RF channel using mainly the muonic component of the EAS. This paper details the 2023 configuration of the BEACON prototype, covering the instrument upgrades in section \ref{sec:instr}, the instrument calibration in section \ref{sec:calibration}, a CR search analysis in section \ref{sec:analysis}, hybrid-instrument performance verification from simulations in section \ref{sec:sim}, and a characterization of the resulting candidate CR events in section \ref{sec:cr}.



\section{\label{sec:instr}The BEACON prototype instrument upgrade in 2023}

\begin{figure*}[!t]
    \centering
    \includegraphics[width=0.98\linewidth,trim={1.8cm 0cm 3.5cm 0cm},clip]{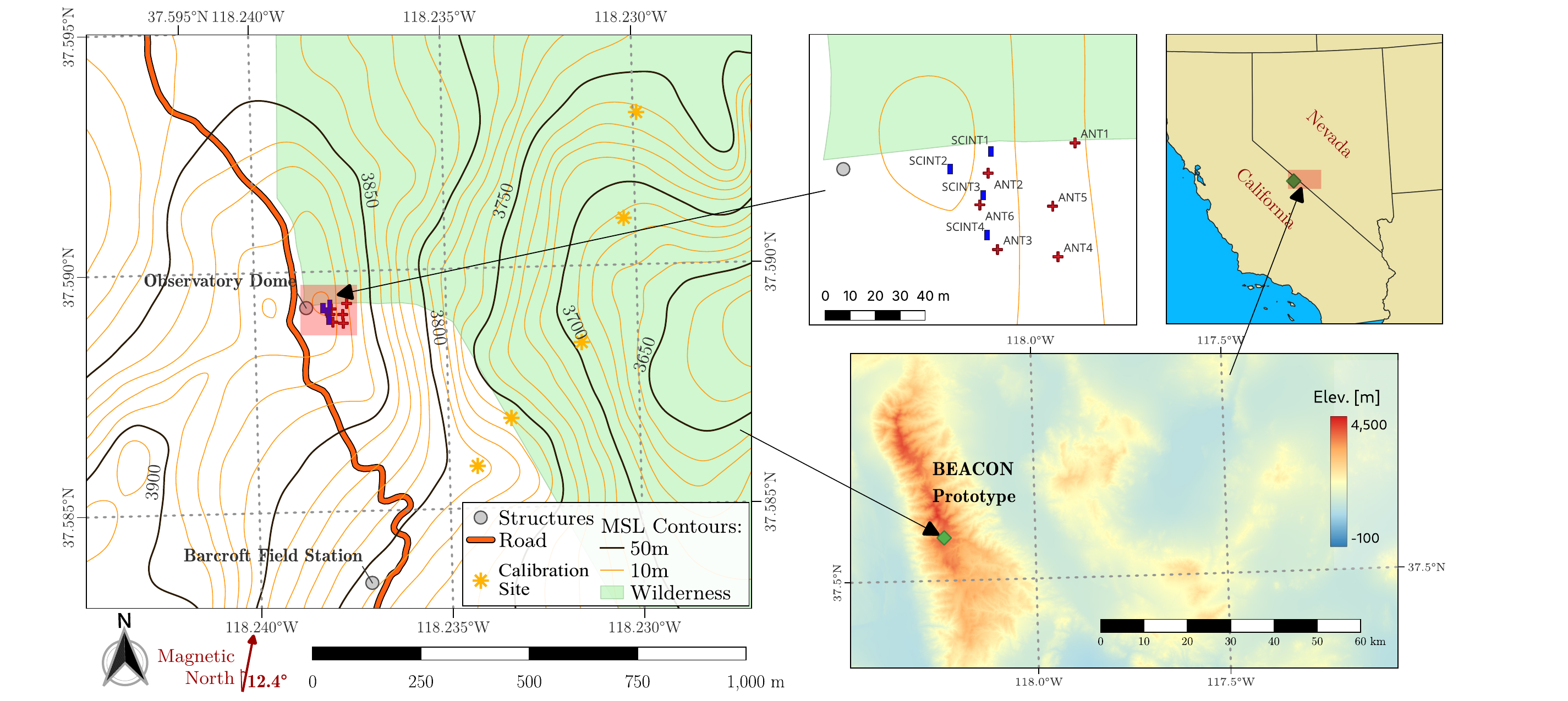}
    \caption{\label{fig:geography}Geographic overview of the 2023 BEACON prototype instrument. The left panel shows the immediate surroundings of the instrument. Contour lines illustrate the local terrain, and a scale bar indicates distance. The right-hand panels include a closer view of the layouts of the RF antenna and scintillator arrays, and two zoomed-out maps providing the broader geographic context, one of which highlights the elevation profile east of the instrument.}
\end{figure*}

The 2023 configuration of the BEACON prototype replaces the four original RF antennas with six newly designed RF antennas, and introduces an independent scintillator array for muon-based validation of the RF signals to improve background estimation. We place four of the RF antennas at the positions of the original four, and the additional two in between. We place four scintillators at the top of the mountain slope, among the westward RF antennas. We continue to use the observatory at the WMRC site as the electronics hub housing the power, network, and data acquisition system (DAQ). The geographic map marking the observatory and the final locations of the detectors along the site is shown in figure \ref{fig:geography}.

\subsection{\label{sec:antennas}Radio antennas and signal chain}

\begin{figure*}[!t]
    \centering
    \includegraphics[width=0.78\linewidth]{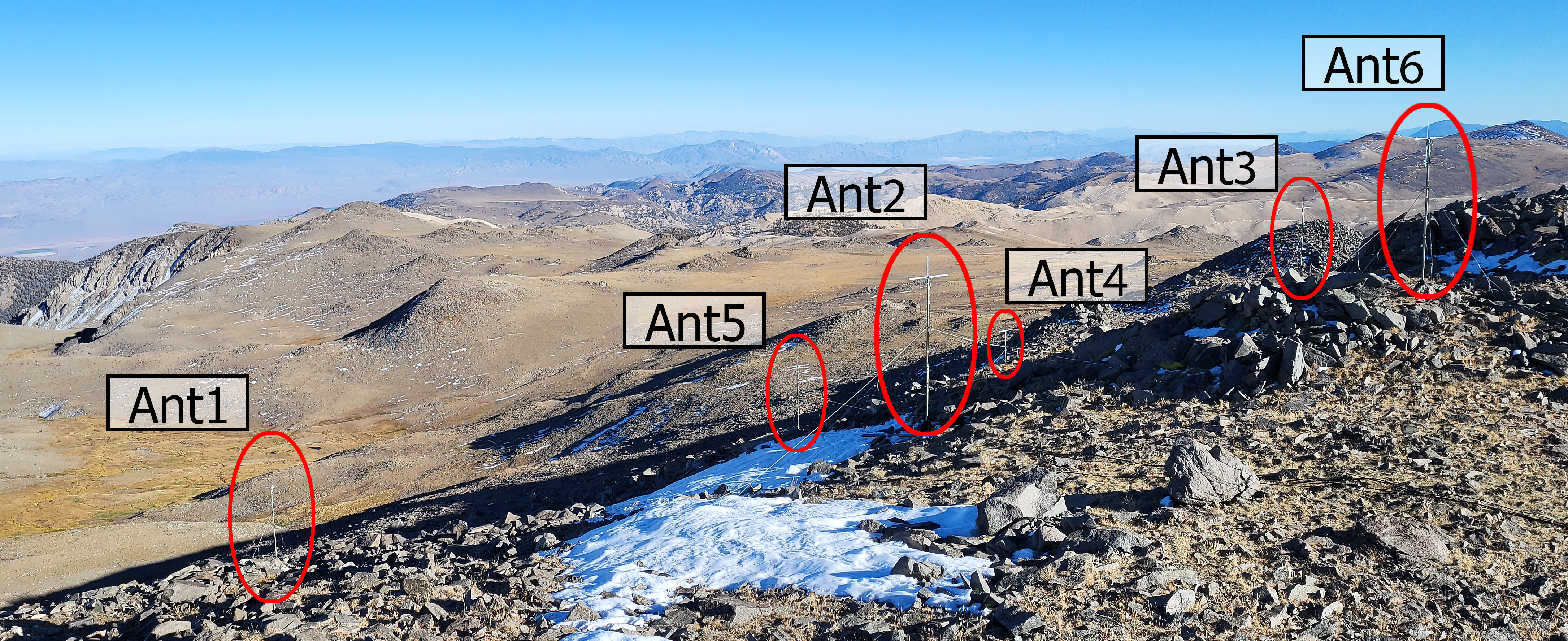}
    \includegraphics[width=0.16\linewidth]{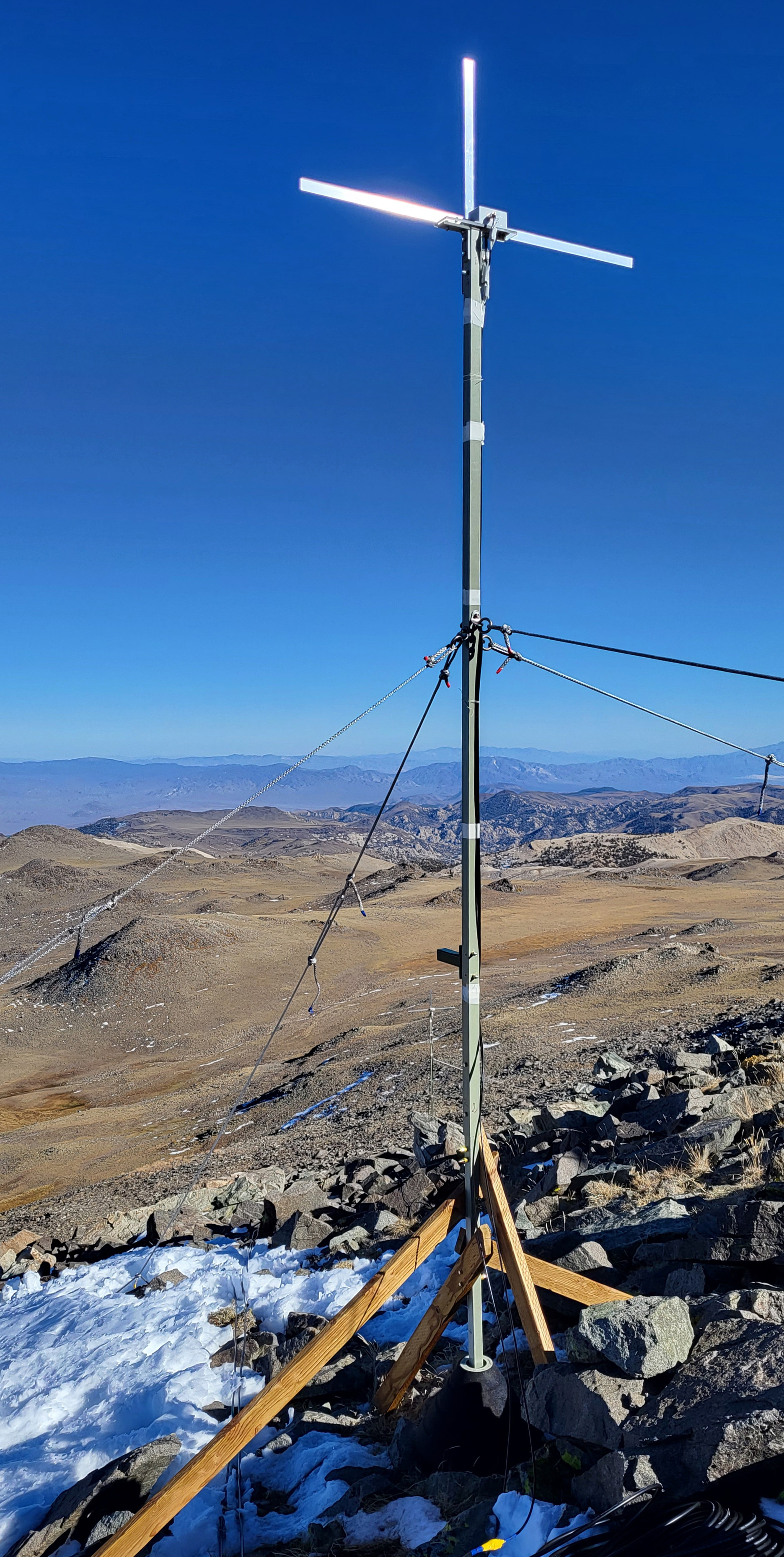}
    \caption{\label{fig:rfarray}Left: The 2023 BEACON prototype RF antenna array overlooking Fish Lake Valley to the east. Right: A close-up of one of the installed crossed-dipole antennas mounted $\sim 3.96$\,m above the ground on a fiberglass mast, supported by a 50\,lb rubber base, wooden struts, and Dyneema guy lines tied to nearby boulders.}
\end{figure*}

The 2023 upgrade replaces the four 0.76 m cylindrical tine antennas with six 0.66 m T-bar tine antennas secured by long acrylic braces. It also replaces the steel mast design holding the antennas roughly 3.96 m above ground with a fiberglass equivalent, removing coupling effects with the conductive material while also improving resistance against twisting and bending. Each antenna is a crossed-dipole design, with the horizontal dipole aligned with the geomagnetic north-south and the vertical dipole aligned up-down, according to the site's east-north-up (ENU) coordinates. The installation of the antennas in 2023 is shown in figure \ref{fig:rfarray}.

After returning to the site the following summer, we found that these changes were successful in preventing the dipole-shearing and mast-warping damage that was seen with the previous design. However, two of the six antennas, located at the top of the mountain, were damaged: one assembly toppled over, and the other snapped at the attachment point of the wooden struts to the fiberglass. Because these failure modes are easier to mitigate than shearing or warping through additional reinforcement at snapping points and more careful securing of the guy lines, we consider the new design to be an overall improvement.

\begin{figure*}[!t]
    \includegraphics[width=0.48\linewidth,trim={1cm 1cm 1cm 2cm},clip]{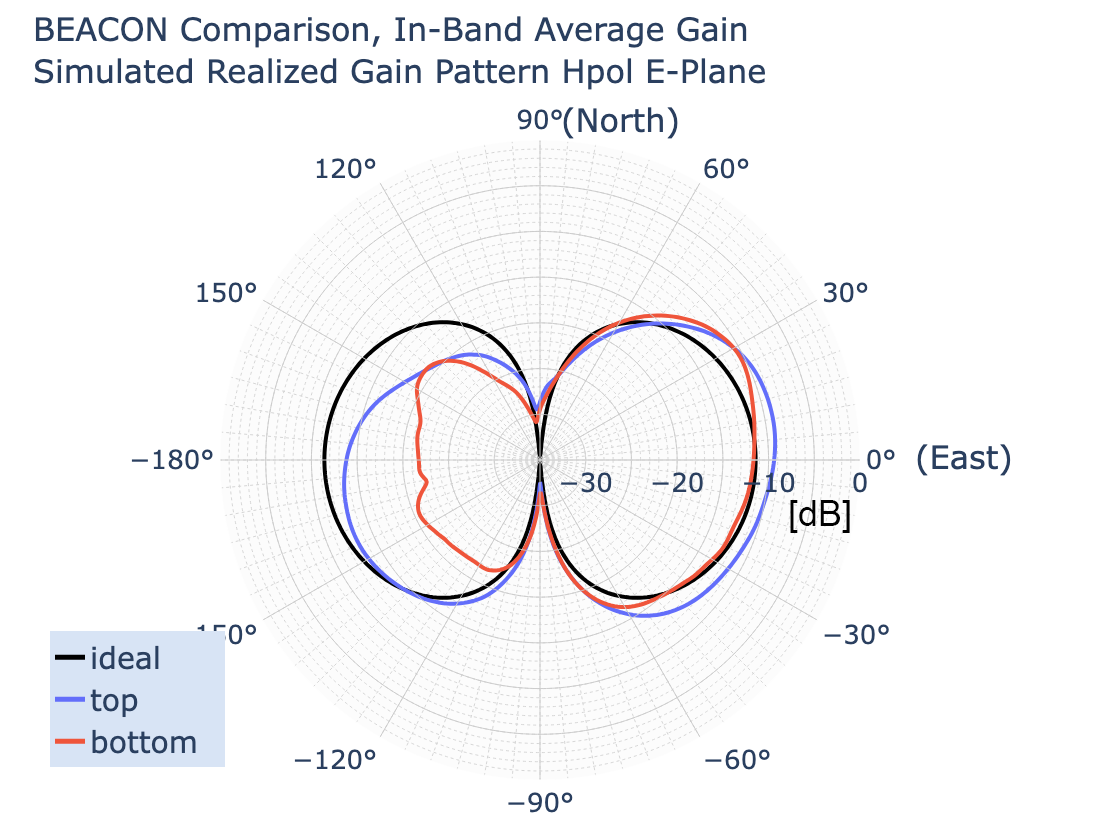}
    \includegraphics[width=0.48\linewidth,trim={1cm 1cm 1cm 2cm},clip]{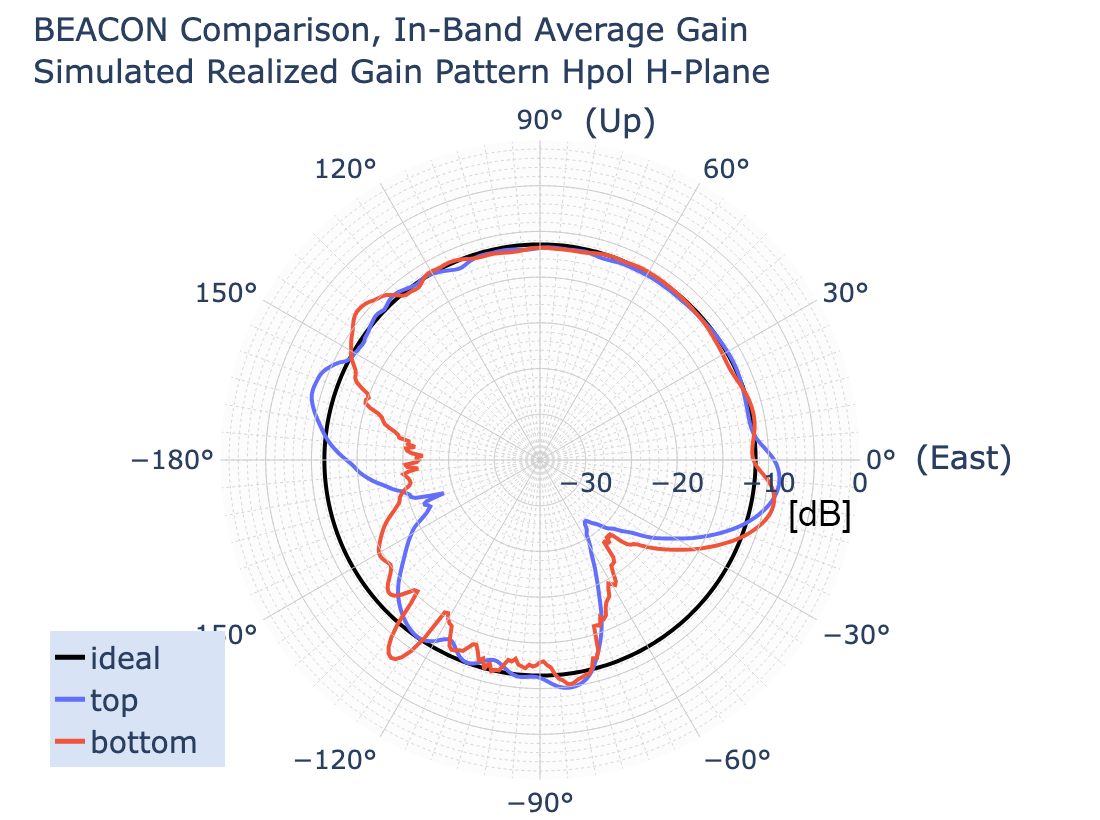}

    \medskip
    
    \includegraphics[width=0.48\linewidth,trim={1cm 1cm 1cm 2cm},clip]{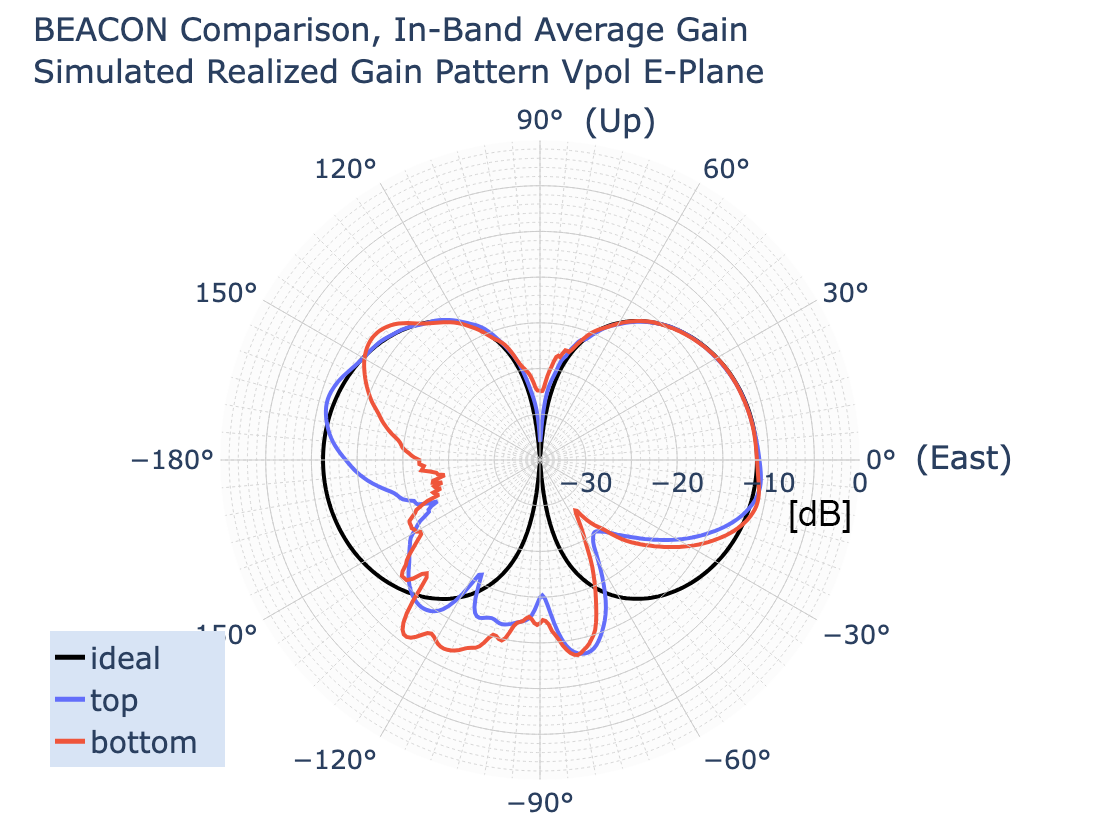}
    \includegraphics[width=0.48\linewidth,trim={1cm 1cm 1cm 2cm},clip]{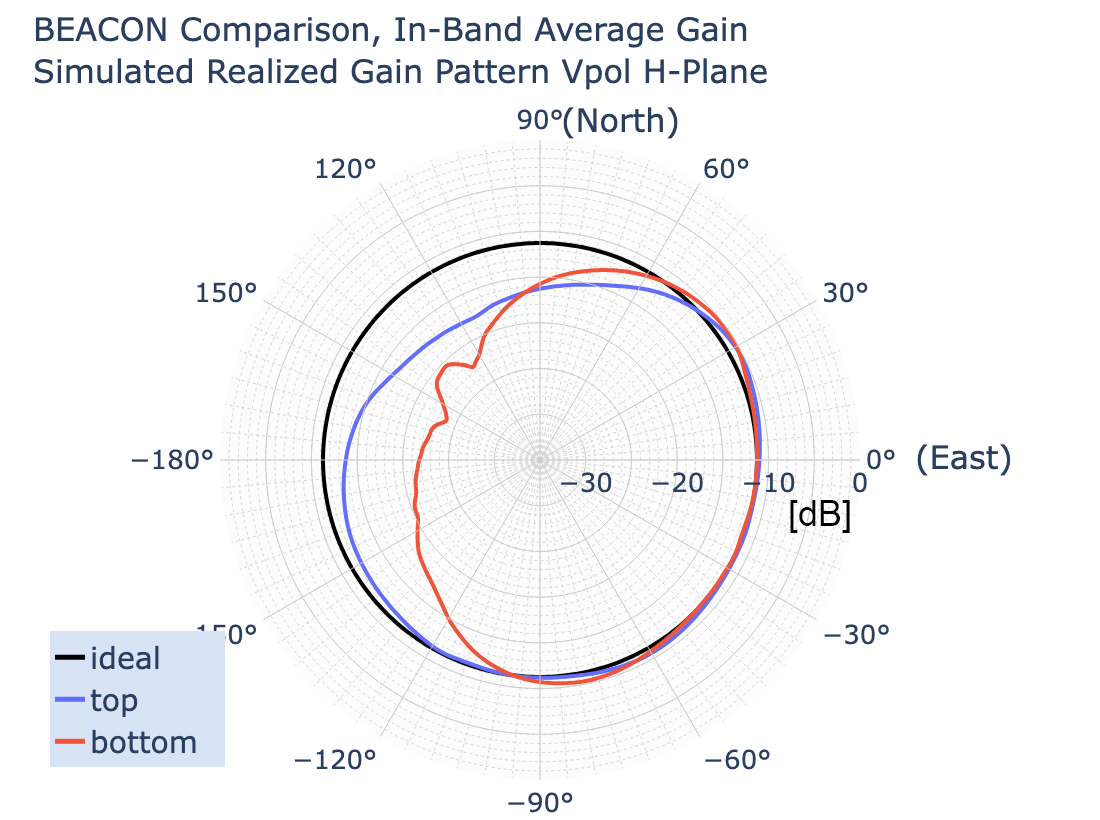}
    \caption{\label{fig:patternslices}The realized gain patterns along the principal planes of the XFdtd-simulated antenna responses averaged over BEACON's 30 -- 80 MHz band. The three curves compare the ``ideal'' (groundless) response to the ground-simulated responses of antennas at the ``top'' and ``bottom'' of the array along the mountain slope. The ground-simulated response features seen on the Hpol E-plane and Vpol H-plane arise from the 10$^\circ$ north-south tilt of the mountain; features in the Hpol H-plane and Vpol E-plane arise from the flat ground west of the array, and the 30$^\circ$ downward slope of the mountain east of the array.}
\end{figure*}

We simulate the electromagnetic (EM) properties of the new antennas in XFdtd \cite{RemcomXFdtd}, a 3D electromagnetic simulation software package, by emitting a pulse from the modeled antenna feeds and deriving their response using the resulting far-field E-fields. We compute the ``normalized antenna response'', $\mathbf{H}_N(f; \theta_\text{a},\phi_\text{a})$,\footnote{We denote $(\theta_\text{a},\phi_\text{a})$ as the spherical coordinates in the antenna's frame used in XFdtd, which aligns the spherical pole with boresight to the east, and denote $(\theta,\phi)$ as the elevation and azimuth coordinates used in analysis.} following the definitions in ref. \cite{farr1998timedomain}. For a voltage excitation, $V_\text{src}$, that produces a radiated E-field, $\mathbf{E}_\text{rad}$, the frequency-domain computation of the antenna response is
\begin{equation}
    \mathbf{{H}}_N(f;\theta_\text{a},\phi_\text{a}) = -i\frac{R\,c}{f}\sqrt{\frac{Z_c}{Z_0}}\frac{\mathbf{E}_\text{rad}(f;\theta_\text{a},\phi_\text{a})}{ V_\text{src}(f)}~,
    \label{eqn:antresp}
\end{equation}
where $R = 1$\,m is set by XFdtd's near- to far-field transformation \cite{balanis-ntff}, $Z_c = 200\,\Omega$ is the approximate impedance at the antenna output, and $Z_0 = \mu_0c \approx 120\pi\,\Omega$ is the characteristic impedance of free-space. The vector components are complex, encoding both the magnitude and phase responses. Both of the channels, H-polarization (Hpol) aligned north-south and V-polarization (Vpol) aligned up-down, have corresponding antenna responses.

We perform three simulation configurations to capture an estimated effect of the ground. The base configuration simulating the ``ideal'' response includes no ground or mast --- just the crossed-dipole. The other two simulations include a simplified ground geometry, using two flat ground planes tilted 10$^\circ$ above north, with an eastward frontal ground plane additionally tilted 30$^\circ$ below east to represent the mountain slope. This geometry simulates the ground effects for an antenna at the ``top'' of the array located at the mountain’s break in slope, where the terrain behind the antenna is flat and the terrain in front follows the mountain slope, as well as for an antenna located at the ``bottom'' of the array, where the mountain slope extends both in front of and behind the antenna.

We execute the simulations using a memory-limited (40 m)$^3$ volume centered on the antenna, thereby simulating distances of up to 20 m from the antenna in any Cartesian coordinate. We compare the simulations' beam patterns quantified by the realized gain,
\begin{equation}
    G(f; \theta_\text{a},\phi_\text{a})~[\text{dB}] = 10\log_{10}\left(\frac{4\pi f^2}{c^2}|\mathbf{H}_N(f; \theta_\text{a},\phi_\text{a})|^2\right) ~.
    \label{eqn:ant_gain}
\end{equation}
The average in-band gain is compared in figure \ref{fig:patternslices}, 
displaying for each polarization channel the resulting principal E- and H-plane slices, which characterize the polarization-aligned and polarization-anti-aligned planes \cite{balanis2016antenna}. Compared to the ``ideal'' dipole, features in the Hpol H-plane and Vpol E-plane patterns are apparent according to the slopes of the ground planes. These patterns also show a significant reduction in gain, as well as azimuthal asymmetry for directions behind (west of) the antenna, as seen in the Hpol E-plane and Vpol H-plane.

\begin{figure*}[!t]
    \centering
    \includegraphics[width=0.95\linewidth]{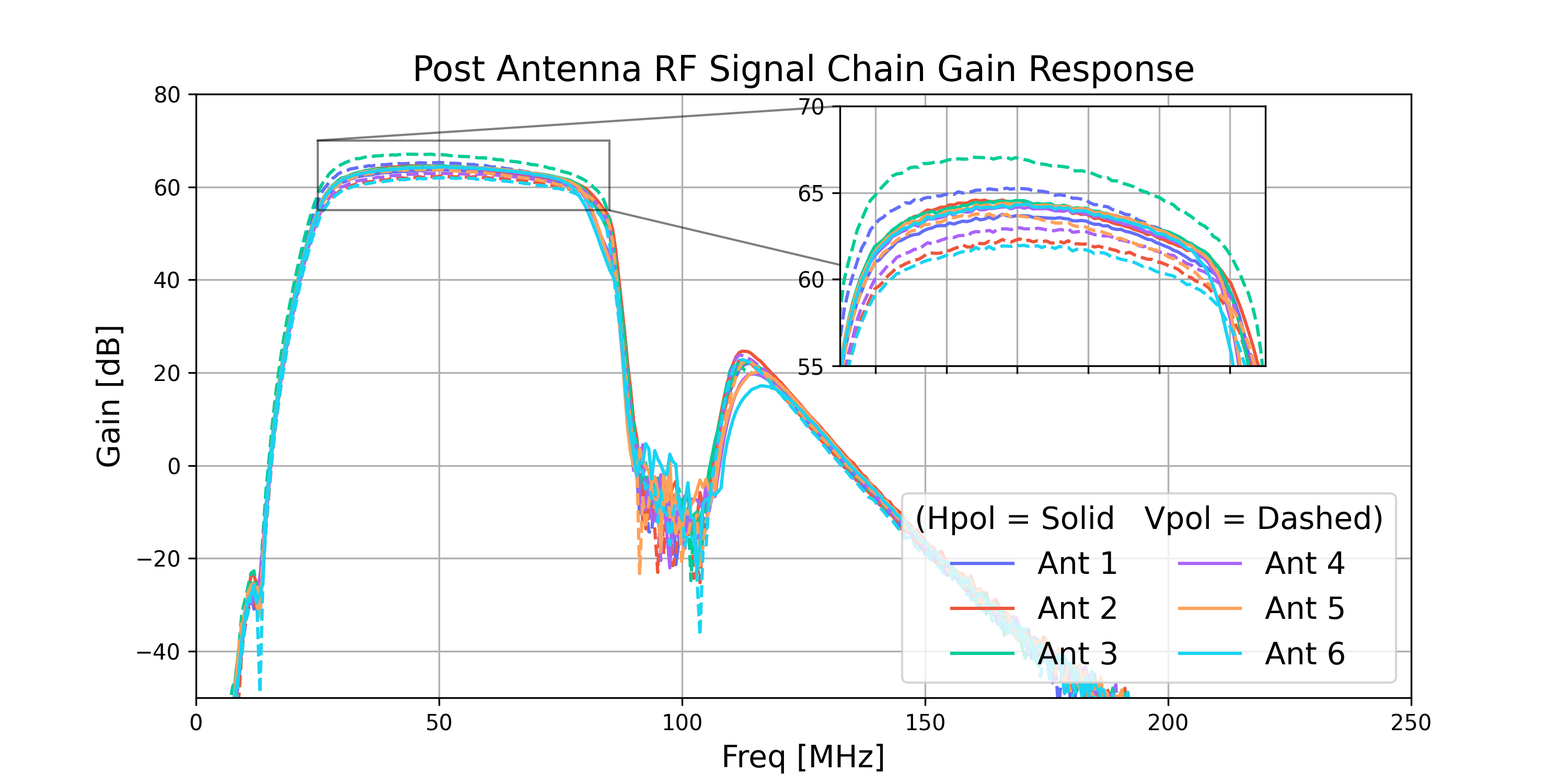}
    \caption{\label{fig:rfchains21}The cascaded gain response of the post-antenna RF signal chain for each channel (labeled by antenna number and polarization). This includes the $\sim$40\,dB antenna-mounted pre-amplifier, cables, lightning arrestor, 88 -- 108 MHz band-stop filter, and the $\sim$29\,dB second-stage RF receiver amplification in the data acquisition system.}
\end{figure*}

A signal received by the antenna passes through an RF signal chain comprising a $\sim$40\,dB antenna-mounted pre-amplifier (AMPA), a lightning surge arrestor, an 88 -- 108 MHz band-stop filter, and about 120 m of coaxial cable. A transformer in the AMPA converts the balanced dipole signal to an unbalanced 50\,$\Omega$ signal. At the DAQ, RF receiver boards provide an additional $\sim 29$\,dB amplification before digitization (see section \ref{sec:daq}).

With a Keysight FieldFox vector network analyzer (VNA), we measure and characterize the AMPA and RF receiver boards in-lab, and the remaining components between the AMPA and the DAQ in situ per channel. The full cascaded RF chain response gain after the antenna is shown in figure \ref{fig:rfchains21}, not including the gain at the digitization step. In total, the RF signal chain amplifies a signal in the 30 -- 80 MHz band by $\sim 63$\,dB.

\subsection{\label{sec:scint}Scintillators}

\begin{figure*}[!t]
    \centering
    \includegraphics[width=0.51\linewidth]{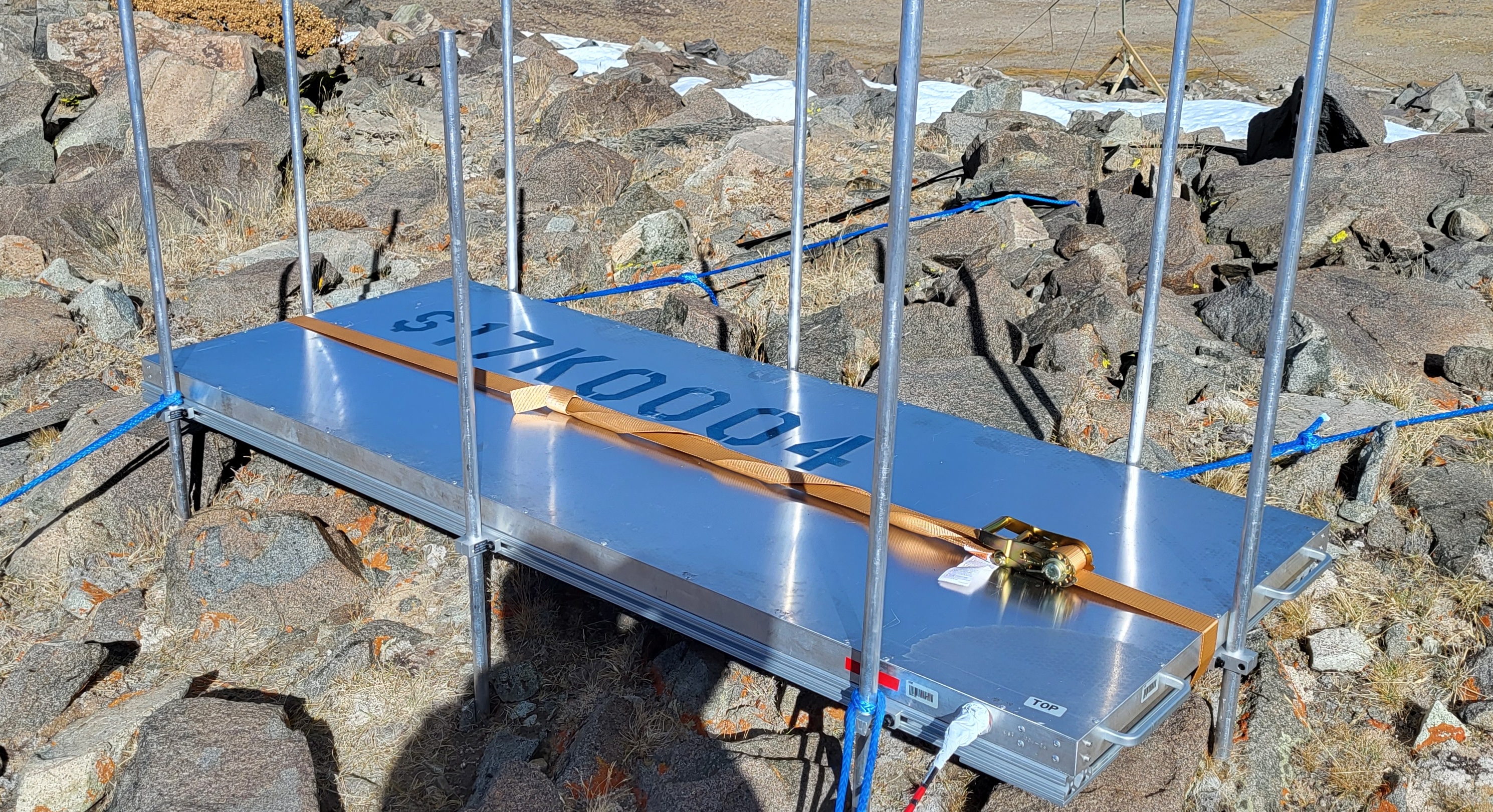}
    \includegraphics[width=0.467\linewidth]{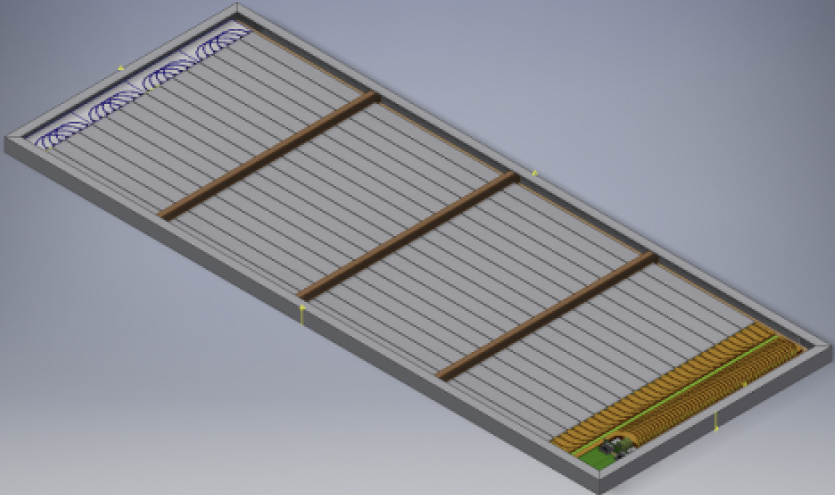}
    \caption{\label{fig:scint}Left: A scintillator detector mounted in situ on a custom frame above the rocks. Right: A digital model showcasing the inner 1.5\,m$^2$ scintillation material, adapted from ref. \cite{scint-Huber2021_1000131545}.}
\end{figure*}

The 2023 BEACON prototype includes four IceTop scintillator detectors, produced by the Karlsruhe Institute of Technology \cite{scint-Huber2021_1000131545}, installed at the top of the mountain slope among the westward RF antennas. An installed detector is shown on the left in figure \ref{fig:scint}.

The available cable lengths and the requirement to minimize wind loading constrain the location of each scintillator. The scintillator cables consist of 70 m (60 m for channel 1) of CAT7 cable routed to the DAQ. The detectors require relatively flat ground both for stable installation and to prevent them from sliding down the mountain.

We install the scintillators on a custom 80/20 aluminum T-slot frame with height-adjustable attachments to eight 2.54 cm outer-diameter solid aluminum poles and orient them with a $\sim 15^{\circ}$ tilt eastward, approximately parallel to the ground surface and just above the rocks in order to minimize wind loading. This orientation also directs the face of the scintillator, where its sensitivity is maximized, closer to the expected arrival direction of radio-detectable CRs, as verified in section \ref{sec:sim}.

The inside of each scintillator unit is shown in the digital model on the right in figure \ref{fig:scint}. As described in ref. \cite{scint-Huber2021_1000131545}, each scintillator contains an array of 16 plastic polystyrene-based scintillation bars, each with dimensions 5 cm $\times$ 1 cm $\times$ 1.875 m, giving a total sensitive area of 1.5\,m$^2$. Optical fibers connect all scintillation bars to a silicon photomultiplier (SiPM). The design allows for either a digital or an analog output, and we use the analog output in order to process and digitize the signals ourselves in the BEACON DAQ.

\subsection{\label{sec:daq}Data acquisition system (DAQ)}

\begin{figure*}[!t]
    \centering
    \includegraphics[width=0.95\linewidth, trim={0 0.5cm 0.5cm 0},clip]{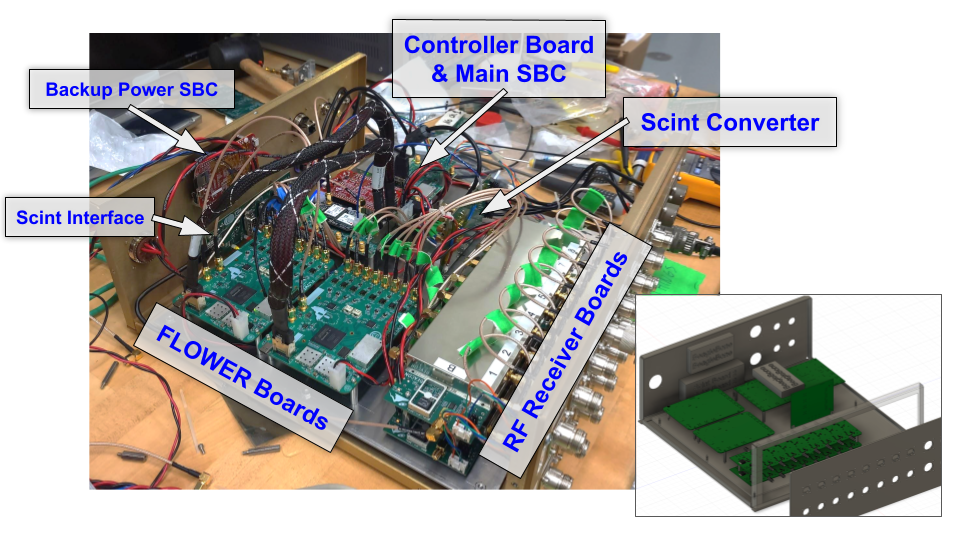}
    \caption{\label{fig:daq}The 2023 BEACON prototype DAQ. The enclosure lid is removed to display the internal subsystems. A digital model is shown at the bottom right to aid visualization of the board layout. An RF signal first passes through the stacked RF receiver boards, while a scintillator signal first passes through the multi-to-single-line converter; both then enter the FLOWER boards for digitization. Details for each DAQ component are discussed in the text.
    }
\end{figure*}

With new instrumentation and more channels needing to be digitized, we install a new DAQ, shown in figure \ref{fig:daq}. A custom controller board routes and regulates power throughout the DAQ subsystems, and is connected to a BeagleBone Black single-board computer (SBC) for commanding. Two FLOWER (FLexible Octal WavEform Recorder) boards~\cite{agarwal2025instrumentdesignperformanceseven} use an analog-to-digital converter (ADC) chip to digitize signals with a 1\,V range across 255 analog-to-digital units (ADU), and comprise a total of 16 channels at 250\,MSa/s each, which can be configured to perform beamforming or coincidence triggering on a selected channel subset. The SBC handles commanding and readout for the FLOWER boards. Additionally, two 8-channel RF amplification boards provide second-stage amplification of the received antenna signals by $\sim 29$\,dB.

We connect the scintillators through a custom scintillator interface board, which converts the multi-line input to a single-line output, for compatibility with the RF feeds on the FLOWER boards. The interface board also handles communication between the scintillators and the controller board. Commands include powering and controlling the scintillator SiPM gains.

The DAQ draws $\sim 25$\,W of power. We use a second BeagleBone Black SBC to control a DAQ power supply connected to the Barcroft Observatory power system, which allows remote hard resets. Both SBCs are connected to the Barcroft station network, operating via a microwave connection to Bishop from White Mountain. The system copies data over the network to a server located inside Barcroft Station for staging.

While the DAQ is capable of real-time beamforming to produce an RF trigger, we implement only the scintillator-coincidence trigger for the analysis discussed in section \ref{sec:analysis} due to requiring further tests of the firmware for the new DAQ. However, a preliminary implementation of the RF trigger would recreate the beamforming trigger logic in the previous prototype.



\section{\label{sec:calibration}Calibration}

To correctly reconstruct properties of a CR EAS from an observed event, we calibrate the detector positions and signal timing for both the RF and scintillator arrays. We additionally use the calibration data collected by the scintillators to calibrate each of their energy responses.

\subsection{\label{sec:fieldcal}Radio array calibration}

\begin{figure*}[!t]
    \centering
    \includegraphics[width=1.0\linewidth]{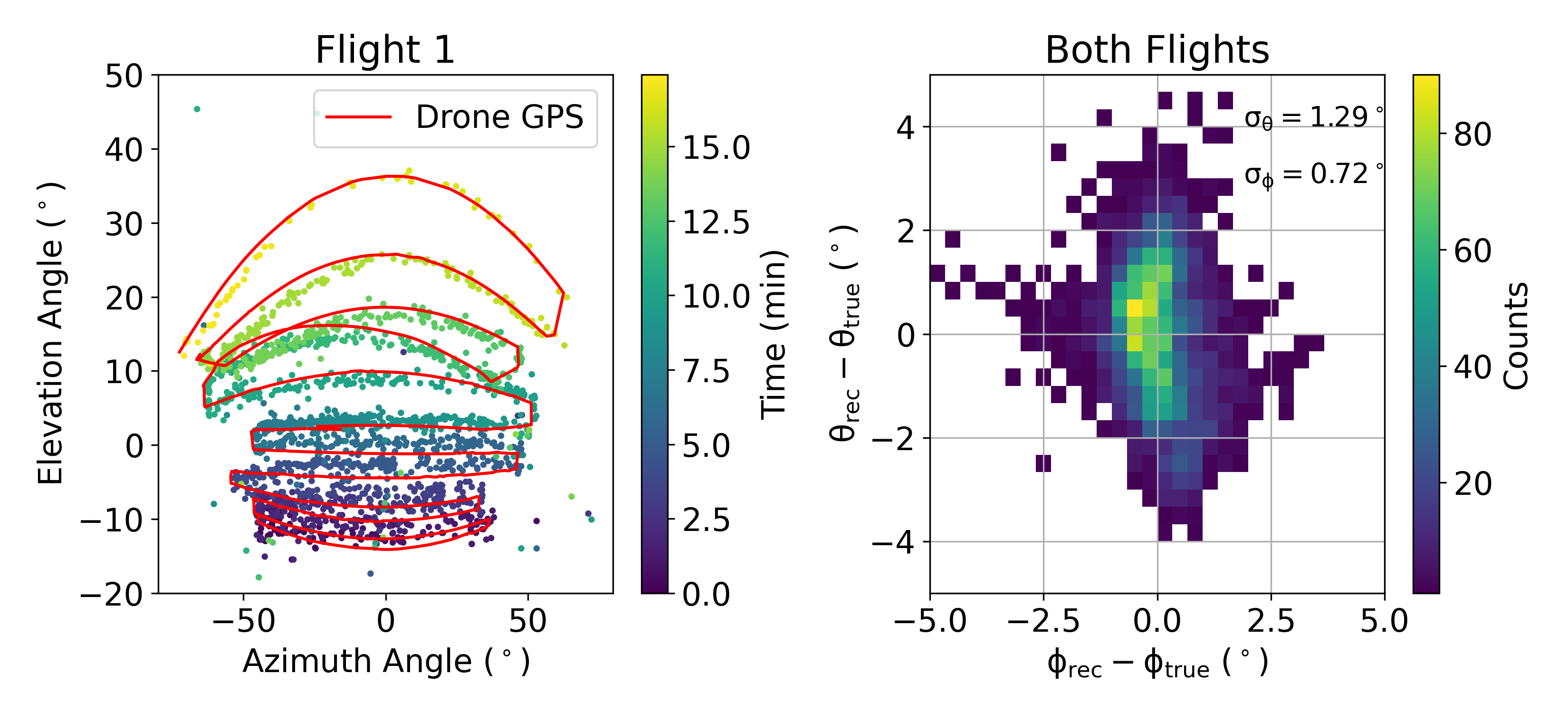}
    \caption{\label{fig:cal}Left: The GPS-recorded flight path of the drone overlaid on the reconstructed locations of radio pulses recorded during the drone flight. Right: A 2D distribution of the reconstructed direction errors in elevation and azimuth combining the two drone flights. Here, we subtract a nonzero mean error caused by a systematic offset in the drone GPS.}
\end{figure*}

We measure the physical positions of the six RF antennas with real-time kinematic (RTK) positioning (nominal cm-precision) at $\sim 1.52$\,m above-ground-level (AGL) on the mast and add $\sim 2.44$\,m based on pre-installation mast-length measurements, to report the height of the antenna center.

We calibrate the RF antenna phase center positions to minimize residuals on the reconstructed direction for pulses transmitted from known locations via ground and drone pulsing. The ground pulsing consists of five sites that are $\sim 500$ -- 900\,m from the BEACON array, spanning $\sim -60^\circ$ to 60$^\circ$ in azimuth and lying below the horizon in elevation. We measure the location of each pulsing site with RTK positioning. Three of the five sites utilize a broadband biconical antenna, while the other two use a BEACON dipole antenna. A FID high-voltage pulse generator excites the ground-calibration antennas, emitting at a constant GPS-synchronized rate, so that a fixed-rate software trigger on the DAQ can consistently capture the pulse.

We calculate the propagation time along each antenna cable using in-band group delays measured with a VNA. Using these times as an initial estimate, we determine the antenna phase center positions and cable lengths by minimizing the error on the reconstructed source direction of each calibration event from the known position of the source. This minimization process is detailed in references \cite{beacon2021-Southall_2023, danthesis-2022PhDT19S, andrewthesis-Zeolla:2026pan}.

We perform the minimization process using only the RF Hpol channels. The RF Vpol channels are distributed across the two FLOWER boards on the DAQ, and a timing synchronization issue between the boards introduces significant variance in the relative timing of the split channels. For this reason, maps of reconstructed direction for RF throughout this work are constructed from Hpol channels.

For drone pulsing, we use a DJI Matrice 300 RTK drone, which generates impulsive radio noise at a high rate during regular operation. Due to the irregular intervals between pulses and an observed drift in the timing recorded by the drone, we elected to only utilize the ground pulsing for calibration and to use the drone flights for validation. 

Over two drone flights, we covered a range of $\sim -60^\circ$ to 60$^\circ$ in azimuth and $\sim -10^\circ$ to 50$^\circ$ in elevation. The path and reconstructed directions of the first drone flight are shown on the left in figure \ref{fig:cal}. The right of the figure shows the 2D distribution of errors in reconstructed elevation and azimuth for both flights. We observe an offset in azimuth in flight 1 ($\mu_\phi=1.53^\circ$), but not in flight 2 ($\mu_\phi=0.0^\circ$). Similarly, we observe a larger offset in elevation in flight 2 ($\mu_\theta=-1.42^\circ$) than in flight 1 ($\mu_\theta=-0.47^\circ$). We also find that the angular error in both flights increased as the drone approached the array. We attribute this behavior to a constant systematic offset in the recorded GPS position of the drone, which may change when the drone is power-cycled between flights. In the figure, we subtract these offsets so that we can directly observe the spread in the pointing error. We observe a standard deviation of $1.29^\circ$ for the reconstructed angular resolution in elevation, and $0.72^\circ$ in azimuth.

\subsection{\label{sec:scintcal}Scintillator array calibration}

\begin{figure*}[!t]
    \centering
    \includegraphics[width=0.95\linewidth]{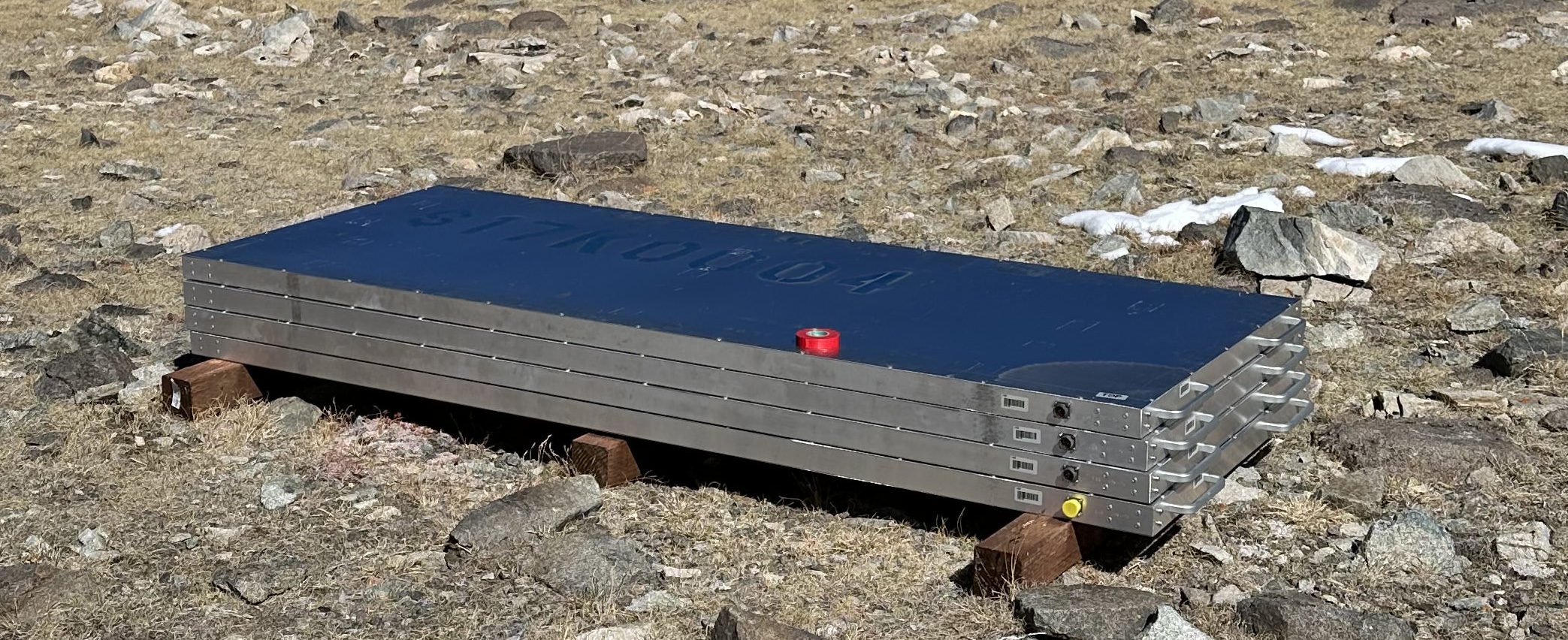}
    \includegraphics[width=0.95\linewidth]{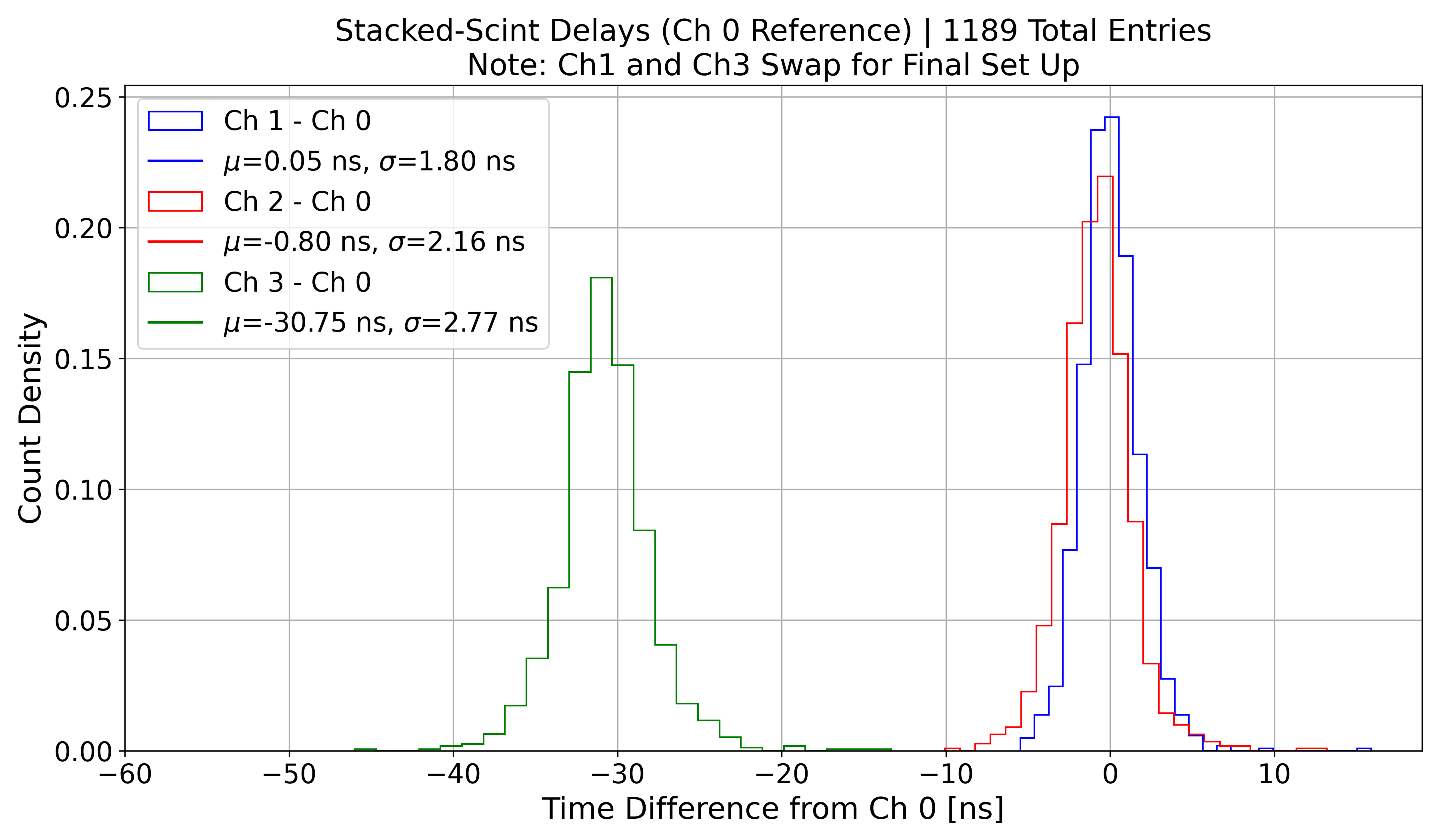}
    \caption{\label{fig:scintdelays}Scintillator timing delays between signals observed during a stacked scintillator configuration, referenced to channel 0 (the topmost scintillator). The $\sim 30$\,ns offset seen on channel 3 is due to its cable being $\sim 10$\,m shorter than those used on channels 0 through 2. We use these measurements to correct the timing of signals seen by each scintillator in the array.}
\end{figure*}

We also calibrate the scintillator positions using RTK positioning, measuring the scintillator panel corners and using the average as the center. To calibrate the signal timing and energy response of each scintillator, we use data collected with the stacked scintillator configuration shown at the top of figure \ref{fig:scintdelays}.

We determine timing delays from differences in the rising-edge arrival times measured in the stacked dataset after discarding events containing low-amplitude, near-maximum-amplitude, and multiple peaks to ensure a clean calibration set. We measure arrival times using a 50\% constant-fraction-discriminator threshold on the pulse amplitude along the rising edge, which we linearly fit between the 20\% and 80\% amplitude points, and reference all arrival times to channel 0, corresponding to the topmost scintillator in the stack.

The distribution of time delays is shown at the bottom of figure \ref{fig:scintdelays}. The near-zero time delays for channels 1 and 2 indicate agreement with the timing of channel 0, whereas the time delay for channel 3 shows a $\sim 30$ ns offset from channel 0. This is due to a $\sim 10$ m shorter cable being used on channel 3 during the stacked configuration (it is used on channel 1 in the final configuration). When the scintillators are used to reconstruct the direction of the source, we use these delays as references for the propagation times. Additional offset to these is due to differences in arrival time based on the array geometry.

We calibrate each scintillator's energy response using a method similar to that described in ref. \cite{Thoudam_2014}. Using data collected with the stacked scintillator configuration, we identify the integrated ADU-$\mu$s signal distribution produced by through-going minimum-ionizing muons. We fit a Landau-Gaussian convolution to this distribution and take the peak of the fit as the mean signal corresponding to a single minimum-ionizing particle (MIP). These results are shown in table \ref{tab:scint_MIP}.

\begin{table}[t]
    \centering
    \begin{tabular}{c@{\hspace{1.5cm}}c}
        \toprule \toprule
        Scintillator \# & $S_{\mathrm{MIP}}$ [ADU-$\mu$s] \\
        \midrule
        1 & $6.106 \pm 0.003$ \\
        2 & $6.098 \pm 0.006$ \\
        3 & $6.105 \pm 0.004$ \\
        4 & $5.451 \pm 0.005$ \\
        \bottomrule
    \end{tabular}
    \caption{Peak values in ADU-$\mu$s of the measured integrated signal distributions for each scintillator during the 2023 stacked-configuration data runs.}
    \label{tab:scint_MIP}
\end{table}

To determine the energy deposited into the scintillators from such events, we conduct simulations in Geant4 \cite{geant4} using a model of the scintillators following the description in ref. \cite{scint-Huber2021_1000131545}. We initiate single 4 GeV muons at random points on the model surface with uniformly sampled azimuth angles and elevation angles sampled from a $\sin\theta\cos^3\theta$ distribution arising from the expected $\cos^2\theta$ distribution of muons in the flux and a $\cos\theta \sin\theta$ geometric factor for a flat detector \cite{Thoudam_2014}. We take the most probable value in the distribution of 100,000 simulated events as the mean energy deposition, and obtain $1.856 \pm 0.004$ MeV. The mean energy of muons at sea level is 4 GeV \cite{AUTRAN201877}, but given the significantly higher elevation at the BEACON prototype site, the value may differ in reality. To account for this, we repeat the simulation process for muon energies 1 -- 10\,GeV where the muon stopping power is expected to be approximately constant, and we find a maximum variation of 2\% in the energy depositions.

Assuming a linear relationship, we obtain the deposited energy, $E^\text{sc}_i$, into scintillator $i$ for a given event from the integrated signal, $\mathrm{S}_i$, as
\begin{equation}
    \label{eq:scint_energy_conversion}
    E^\text{sc}_i = (1.856\,\mathrm{MeV})~\frac{\mathrm{S}_i\,[\text{ADU-}\mu\text{s}]}{S_{\mathrm{MIP},i}}~,
\end{equation}
where each $S_{\mathrm{MIP},i}$ corresponds to the signal value in table \ref{tab:scint_MIP}.



\section{\label{sec:analysis}Data analysis}

We present a CR search analysis with the 2023 BEACON prototype targeting the identification of candidate CR events that have a detectable radio signal among a scintillator-triggered dataset. The scintillator coincidence (COIN) trigger on the DAQ requires at least three of the four scintillator detectors to reach amplitudes $\geq 5$ ADU. The COIN-triggered dataset comprises 25 million events over approximately 83 days of stable operation between October 19, 2023 and March 2, 2024. A seven-week winter power outage halted data taking within this period, but did not measurably affect detector performance after operations resumed. The end of the dataset corresponds to an RF antenna becoming nonoperational after storm-related damage.

The goal of the analysis is to identify candidate CR events whose radio signals are statistically distinguishable from those of RFI background sources. In order to do this, we build selection criteria, or analysis cuts, based primarily on radio characteristics, which we define in the following sections. We then use these criteria to select events in the COIN-triggered dataset that contain a radio signal whose characteristics are inconsistent with radio signals observed in events triggered using a fixed 2 Hz software (SW) trigger in the DAQ.

Among the SW-triggered dataset, about 95\% of the events consist of thermal noise in the RF channels, defined as events whose coherently summed waveform has an SNR $< 5.3$ (see section \ref{sec:rfsnr}). High-SNR RF events in the other 5\% of the SW-triggered dataset come from site RFI emitted by multiple sources including, but not limited to, anthropogenic noise such as airplanes and electrical infrastructure, and natural phenomena such as lightning and solar activity.

We assume that the COIN-triggered scintillator pulses are produced by CR EAS particles. However, because the scintillators and RF antennas have different energy acceptances and elevation-dependent sensitivities (see section \ref{sec:sim}), low-energy or near-vertical EASs may be observed with a scintillator signal but without an observable radio counterpart. Additionally, a detected radio signal that is coincident with a scintillator signal is most often for this site not associated with an EAS, but rather with RFI originating from the same set of sources detected in the SW-triggered dataset.
 
In section \ref{sec:backgrounds}, we estimate the background of RFI radio signals coincident with scintillator signals by constructing a synthetic dataset of uncorrelated RF-scintillator pairs and applying to it the same criteria used to select candidate CR events.

\subsection{\label{sec:RFchar}Radio characteristics}

RF emission from CR EASs is expected to be a coherent and impulsive signal \cite{emission-Schr_der_2017} that, at the WMRC site, comes from above the horizon predominantly north-south polarized. Cuts built on these properties must effectively eliminate similar RFI backgrounds while preserving real CR events that are correlated with the scintillators.

The radio characteristics used in this analysis are reconstructed direction and the coherently summed waveform's SNR and impulsivity. As noted in section \ref{sec:fieldcal}, we reconstruct directions using only Hpol channels. 

We choose not to use polarization as a cut in this analysis. The expected Vpol signal for CR events is small due to the vertically inclined orientation of the site's magnetic field where the antennas are most sensitive. Because we have a relatively small radio array and the uncertainties on reconstructed polarization have not yet been quantified on observed CR events, we do not expect polarization to reliably discriminate against background RFI during selection. We instead reconstruct polarization only for the remaining candidate CR events and compare to expected CR behavior as a characterization of the instrument's reconstructed polarization uncertainty (see section \ref{sec:pol}).

For each event, we upsample each RF waveform by a factor of 100, subtract its mean, normalize it by its standard deviation, and filter continuous-wave (CW) signal contamination. The CW filtering algorithm follows ref. \cite{Gorham_2018}, which was similarly used in previous BEACON analyses. In brief, the algorithm iteratively identifies high-power frequencies, models their contributions as sinusoids, and subtracts those contributions from the original waveform in the time domain.

\subsubsection{\label{sec:rfrecon}Reconstructed direction}

We reconstruct the signal direction to determine the source location of an event. We compute the normalized cross-correlation value for every antenna pair, across a range of time delays. For each direction on a 0.25$^\circ$ rectangular sky grid, we determine the expected relative time delay for a given antenna pair assuming a far-field plane-wave approximation. We associate each direction with one correlation value for every antenna pair, resulting in pairwise correlation maps.

We then combine the correlation maps using weights, $w$, proportional to the array baseline lengths, since longer baselines provide greater directional sensitivity. This results in the total correlation map,
\begin{equation}
    \mathcal{C}(\theta,\phi) = \frac{\sum_{i>j} w_{ij}\,\bar{c}(v_i,v_j)|_{dt_{ij}(\theta,\phi)}}{\sum_{i>j}w_{ij}}~, \label{eqn:corrmap}
\end{equation}
where $v_i$ and $v_j$ are the waveforms of the same polarization channel of antennas $i$ and $j$, $w_{ij}$ is the corresponding baseline weight between those two antennas, and
\begin{equation}
    \bar{c}(v,u)|_k = \frac{\sum_{l}^{N_\text{samples}} v_l~u_{l-k}}{\sqrt{\sum v^2\sum u^2}}
\end{equation}
is the normalized cross-correlation of the waveforms, $v$ and $u$, evaluated at the index $k$. The evaluation at $dt_{ij}(\theta,\phi)$ in eq. \eqref{eqn:corrmap} indicates the index corresponding to the expected relative time delay for a source at $(\theta,\phi)$. 

We take the direction, $(\theta_\text{max}, \phi_\text{max})$, that maximizes the correlation map as the reconstructed arrival direction of the event. This direction determines the set of expected relative time delays that best align the measured waveforms under the plane-wave approximation. We then use these delays to construct the coherently summed waveform (CSW),
\begin{equation}
    V_\text{CSW}(t) = v_1(t) + \sum^{N_\text{antennas}}_{i=2}v_i(t - t_{i,1}(\theta_\text{max}, \phi_\text{max})),
\end{equation}
where the time delays chosen are relative to antenna 1.

Note that as a result of band limiting in the signal chain, an impulse originally with a single peak can appear as a signal with multiple peaks. The secondary peaks can have amplitudes comparable to the main peak. Signals can therefore correlate strongly not only at the true alignment of their main peaks, but also at offsets where the main peak of one signal aligns with a secondary peak of another. This results in sidelobes in the correlation maps. In some cases, the correlation at a sidelobe exceeds that at the true alignment, causing a misreconstructed direction. Additional antennas reduce this ambiguity. Previous analyses with the 2021 BEACON prototype detail their treatment of sidelobe reconstruction \cite{danthesis-2022PhDT19S, andrewthesis-Zeolla:2026pan}. For this analysis, we do not apply any treatment of sidelobes, however misreconstructed directions are accounted for in the background estimate and are not seen in the correlation maps for the candidate CR events discussed in section \ref{sec:backgrounds}.

\subsubsection{\label{sec:rfsnr}Signal-to-noise ratio (SNR)}

The SNR characterizes the strength of the signal in a waveform. For this analysis, we define it as
\begin{equation}
    \text{SNR} = \frac{V_\text{max} - V_\text{min}}{2\,V_\text{rms}}~,
    \label{eqn:snr}
\end{equation}
where $V_\text{max} - V_\text{min}$ is the peak-to-peak of the waveform. The root-mean-square (RMS) value of the mean-subtracted waveform, $V_\text{rms}$, is equal to its standard deviation and provides an estimate of the thermal-noise level. For an improved estimate of the thermal-noise level, we calculate the RMS for windowed subsets of the full waveform and use the median RMS value as the $V_\text{rms}$ in eq. \eqref{eqn:snr}. The analysis cut is set to select events with CSW SNR $> 5.3$ (see section \ref{sec:backgrounds}), based on previous instrument RF-trigger thresholds and the CSW SNR distribution of events.

\subsubsection{\label{sec:rfimp}Impulsivity}

We define two measurements of impulsivity, $I_1$ and $I_2$, to quantify the waveform signal duration and amount of power within that duration. We then normalize and average the two to obtain the final impulsivity, $I = (I_1 + I_2)~/~2$.

The first measurement of impulsivity, $I_1$, is a scaled average of the cumulative distribution of power. We cumulatively integrate the power from the Hilbert envelope \cite{hilbertenvelope} of the upsampled waveform, beginning at the signal peak, and normalize it by the full cumulative power so that it ranges between 0 and 1. For highly impulsive signals, the cumulative integration should rise quickly near to 1 and then plateau, indicating that most of the power is encapsulated within a short window of the peak and producing an average cumulative power value near 1. For thermal-noise waveforms, the cumulative power linearly grows across the full window, resulting in an average cumulative power near 0.5. To maximally separate thermal noise and high impulsivity, we rescale the average, $\langle \Sigma Pdt\rangle$, to 
\begin{equation}
    I_1 = \frac{\langle\Sigma Pdt\rangle - 0.5}{0.5}~,
\end{equation}
which sets the range to approximately (0,1), where thermal noise is near 0 and highly impulsive signals are near 1. Negative values are possible in waveforms containing subsequent signals of comparable size to the peak within the integration window, however these are typically rare or indicative of a glitch.

The second measurement of impulsivity, $I_2$, is the time duration measured from the same cumulative power distribution, but now for an integration window spanning before and after the signal peak. The cumulative power distribution for thermal noise waveforms is the same --- a linear integration over the full window. For highly impulsive signals, the distribution should begin at a plateau near 0, then quickly rise to near 1, and then plateau again. We extract the time duration between the 20\% and 80\% cumulative power, indicating the approximate rate of change of power (agnostic to the power amplitude). We then input this quantity, $\Delta t$, into an empirically defined function for an impulsivity measure,
\begin{equation}
    I_2 =\left[0.5\,\cos\left(\frac{\Delta t}{78\,\text{ns}}\right)+0.5\right] \exp\left(-\left(\frac{\Delta t}{220\,\text{ns}}\right)^6\right)~,
\end{equation}
which assigns impulsivity to the range $[0,1]$, where $I_2 \gtrsim 0.5$ when $\Delta t \lesssim 120.4$\,ns. The tuned function parameters are determined by distributions of $\Delta t$ calculated from the full dataset to best place the impulsivity values of thermal noise to $I_2 < 0.5$.

Both impulsivity measurements depend on the size of the integration window. To account for any effects dependent on the window size chosen, we calculate the two impulsivity measurements over successively wider windows and take their averages as the representative values.

As with SNR, we use the impulsivity of the CSW for the analysis cut. By construction of the impulsivity, we set the selection at $I > 0.5$ to reject thermal noise (verified in section \ref{sec:backgrounds}).

\subsubsection{\label{sec:rfbg}Spatial masking from RFI clusters}

\begin{figure*}[!t]
    \centering
    \includegraphics[width=0.9\linewidth, trim={2cm 1.1cm 1.1cm 1.5cm}, clip]{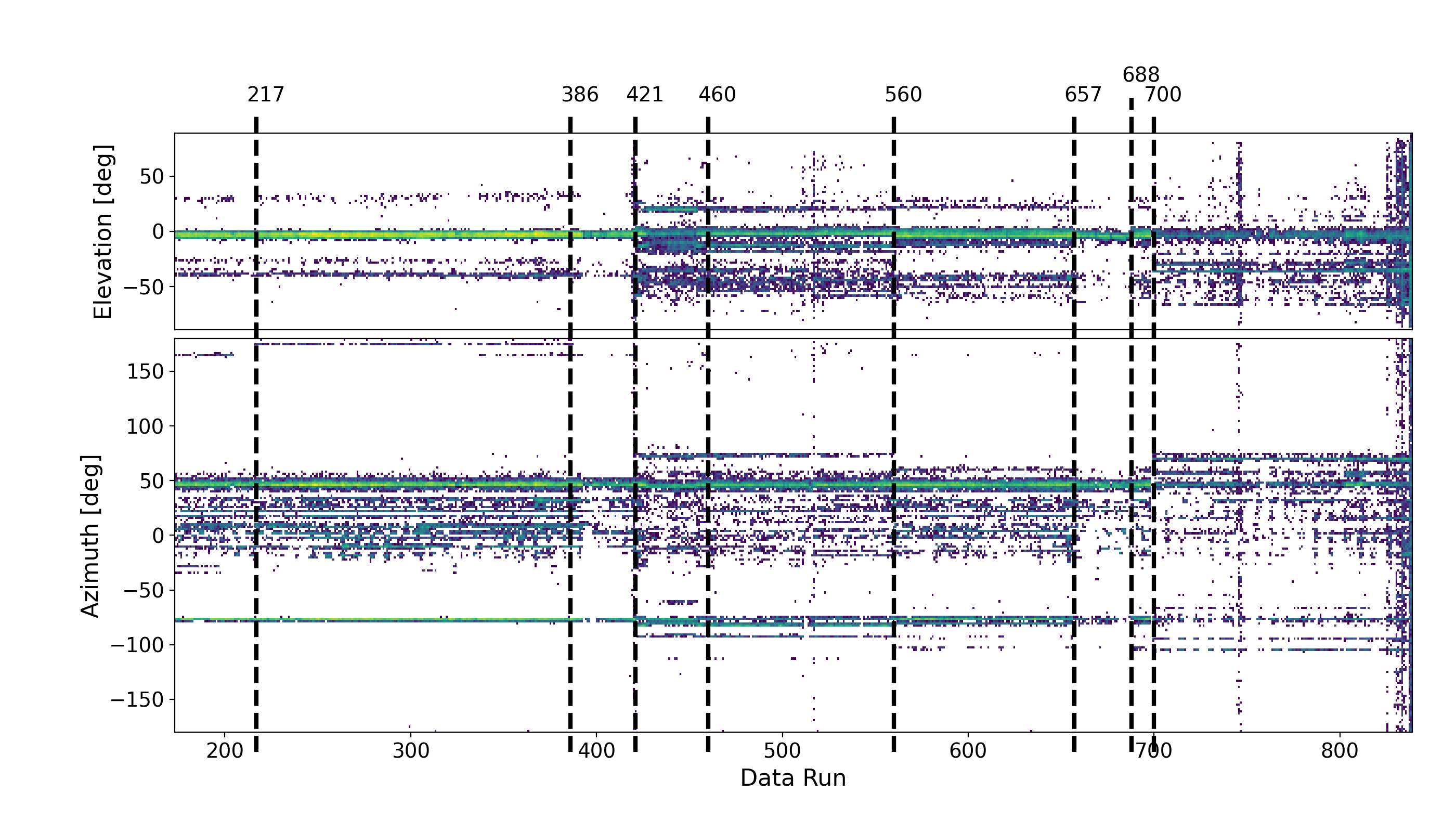}
    \caption{\label{fig:rfitimeperiods}Distributions of reconstructed azimuths and elevations over time from non-thermal RF events in SW-triggered data. Time is represented here as the data run number. Vertical dashed lines separate sudden changes in the reconstructed direction distribution, resulting in nine representative periods of the RFI environment.
    }
\end{figure*}

We separate background RFI events from CR events by characterizing, or scoring, background RFI as spatial clusters in the reconstructed directions of SW-triggered events. We select non-thermal RF events using the SNR cut described in section \ref{sec:rfsnr}. Then we pass the reconstructed directions to the Density-Based Spatial Clustering of Applications with Noise (DBSCAN) clustering algorithm \cite{dbscan-10.5555/3001460.3001507}, implemented through \texttt{scikit-learn} \cite{scikit-learn}, which selects clusters according to a density criterion with a specified distance metric. We specify the metric, $d$, such that for two points with coordinates, $(\theta_i,\phi_i)$, the angular separation is
\begin{equation}
    \cos(d) = \cos(\theta_1)\cos(\theta_2)\cos(\phi_1 - \phi_2) + \sin(\theta_1)\sin(\theta_2)~,
\end{equation}
which accounts for distortions on the rectangular grid on which we define the reconstruction points.

\begin{figure*}[!t]
    \centering
    \includegraphics[width=0.9\linewidth]{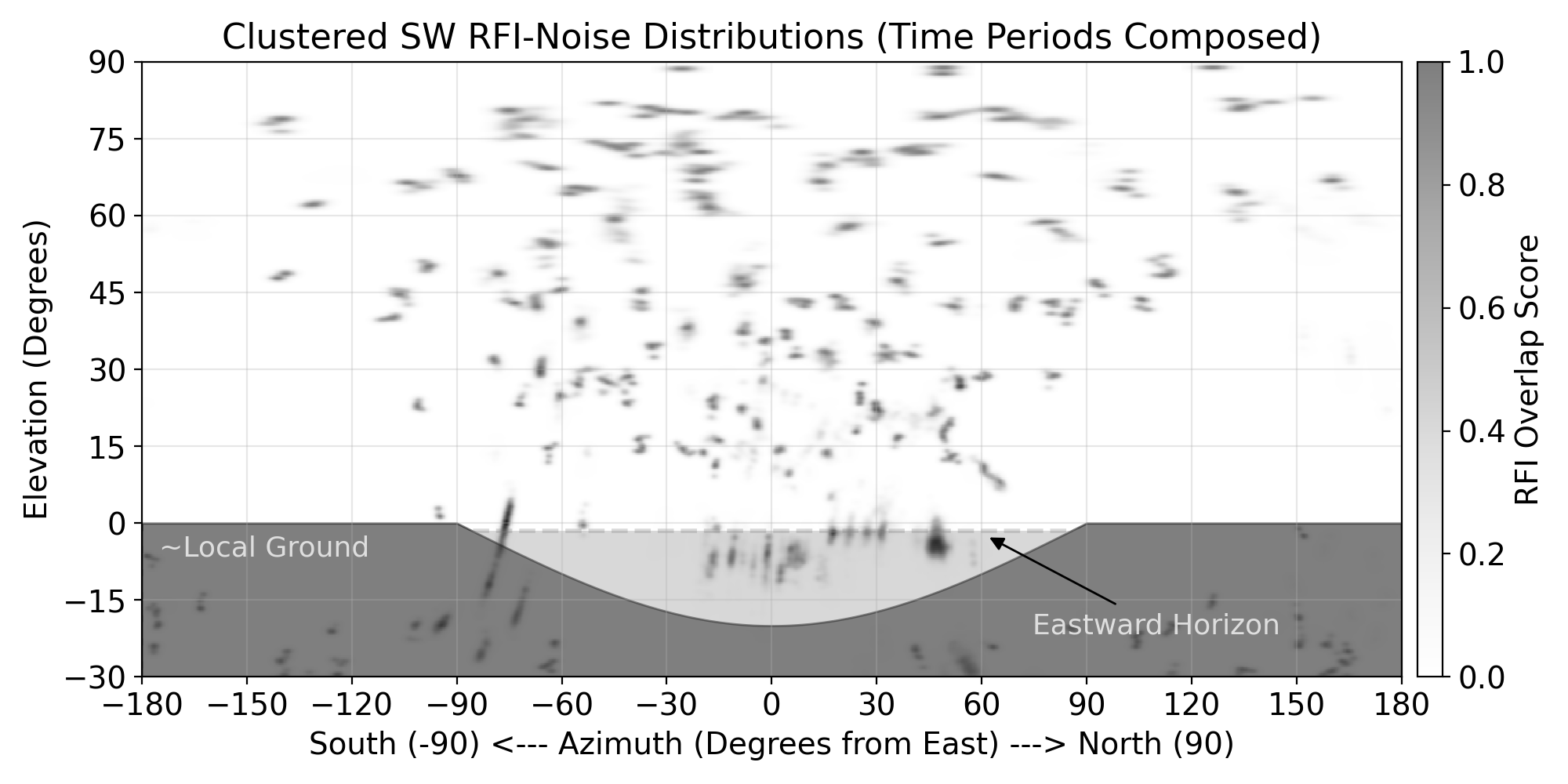}
    \caption{\label{fig:noiseregions}Distribution of RFI across the sky resulting from DBSCAN-identified clusters of impulsive RF events in the SW-triggered dataset. The RFI overlap score quantifies the degree to which a given event is of RFI origin. Clusters identified within each of the nine time periods, shown in figure \ref{fig:rfitimeperiods}, are overlaid on each other. As a result, clusters common to multiple periods will appear darker than the color bar scale's maximum. The approximate local ground of the mountain and the eastward horizon (excluding topography) are indicated. Note that the primary sources of RFI are near the horizon, and many clusters above or below correspond to the reconstruction of their sidelobes.}
\end{figure*}

We optimize the clustering algorithm by sweeping the DBSCAN parameters, \texttt{eps} and \texttt{min\_samples}, which determine the density of any cluster. We choose the final values to identify statistically weaker clusters, which likely correspond to RFI sidelobes, and remove points that DBSCAN flags as noise points, or non-cluster points, which do not meet the density requirements of any nearby clusters.

With each cluster defined by points, we convolve them with Gaussian kernels along azimuth and elevation with widths according to the RF pointing resolution. We normalize each cluster to unity. We call the resulting values RFI overlap scores. From all clusters, this creates a map of RFI overlap scores across the full sky that can be used to define a spatially based data selection cut.

We also account for temporal evolution of the RFI distribution maps. We confine the clustering algorithm within distinct periods of time, separated by sudden changes in the reconstructed direction distribution of SW-triggered events, which appear as persistent bands in azimuth or elevation, shown in figure \ref{fig:rfitimeperiods}. Using the timestamps of these changes as temporal bounds, we identify nine periods of time corresponding to different RFI environments that are relatively stable within each period. These time periods span different lengths, with some spanning one to five days, and others spanning up to 17 or 21 days.

The resulting RFI distribution map is shown in figure \ref{fig:noiseregions}, comprising all the time periods mentioned above. Then, for a given event, we choose the cluster map corresponding to the time period during which the event occurs, and then assign the value on the map at the event's reconstructed direction as the event's RFI overlap score. A score of zero represents an event that is far from any identified RFI cluster, and a score of one corresponds to an event reconstructed close to the core of an identified RFI cluster.

The dominant source of RFI identified in the distribution map is anthropogenic, originating from infrastructure found nearly at the horizon \cite{beacon2021-Southall_2023}. Clusters above the horizon can result from misreconstructed sidelobes of these near-horizon RFI sources.

The clustering method does not identify track-like sources of RFI, such as airplanes, which have been found to be the dominant source of RFI above the horizon \cite{beacon2021-Southall_2023}. We treat airplane accidental coincidences separately, as discussed in section \ref{sec:airplane}, and perform a cross-check with present airplanes on events that pass selection cuts.

\subsection{\label{sec:SCchar}Scintillator characteristics}

The scintillators produce characteristic signals with $\sim 50$ ns rising edges and $\sim 200$ ns exponential decays. We use this behavior to define the scintillator characterization variables used in this analysis.

\subsubsection{\label{sec:screcon}Reconstructed direction}

We use the characteristic rising edge to determine signal arrival times and compare them with the arrival times expected from a modeled particle wavefront to reconstruct the direction.

We treat the particle wavefront as a far-field plane wave. We estimate the uncertainty associated with this approximation as follows. Various models quantify the particle arrival-time distribution as a function of distance from the EAS core \cite{particleArrivalTimeDist-AGNETTA1997301, particlearrivaltimedist-JLinsley_1986, particleArrivalTimeDistribution-refId0}. The relative arrival-time difference increases with distance from the core. At distances greater than 150 m from the core, the time differences can reach 10 -- 20 ns for detectors with a 50 m baseline. This corresponds to an apparent angular offset of up to $\arcsin(c\,\delta t/50~\mathrm{m}) \approx 3.5^\circ$ -- $7^\circ$, representing an upper bound on the pointing error.

For a planar particle front, we compute the expected time delays for each direction on a 0.25$^\circ$ rectangular sky grid. The best-fit direction minimizes the $\chi(\theta,\phi)^2$ map,
\begin{equation}
    \chi(\theta,\phi)^2 = \sum_{i\neq j}^{N_\text{sc}} \frac{\left[\Delta t_{ij}^{\text{meas}} - \Delta t_{ij}^{\text{exp}}(\theta, \phi)\right]^2}{\delta t_{ij}^2}~,
    \label{eqn:sc_recon}
\end{equation}
or equivalently maximizes
\begin{equation}
    \mathcal{P}(\theta,\phi) \propto \exp{(-\chi(\theta,\phi)^2~/~2)}~.
    \label{eqn:sc_reconprob}
\end{equation}
Here, $N_\text{sc}$ is the number of triggered scintillators. We obtain each $\Delta t_{ij}^{\text{meas}}$ by applying a 50\% constant-fraction-discriminator threshold to linear fits of the rising edges of the scintillator pulses, with corresponding timing uncertainties, $\delta t_{ij}$.

We weight the resulting $\mathcal{P}(\theta,\phi)$ maps by the scintillator effective area,
\begin{equation}
    \mathcal{W}^\text{A}(\theta,\phi) = \sum_{j=1}^3\left(A_{j}\,\sum_{i=1}^{N_\text{sc}}\max(0,\mathbf{n_{ij}}\cdot\mathbf{n_0}(\theta, \phi))\right)~,
\end{equation}
where $A_j$ is the area associated with the $j$th scintillator face, $\mathbf{n}_{ij}$ is the unit normal vector for the $i$th scintillator, and $\mathbf{n}_0(\theta,\phi)$ is the unit vector pointing toward $(\theta,\phi)$. This improves the reconstructed direction and sets negative projections to zero by suppressing solutions reconstructed toward the ground beneath the scintillators that arise from the front-back symmetry. We correct for the solid-angle distortion associated with the rectangular sky grid by multiplying this by $\cos(\theta)$. Then, we normalize the resulting weight to a maximum of unity, multiply it by eq. \eqref{eqn:sc_reconprob}, normalize the resulting map to unit sum, and take the direction corresponding to the map maximum as the reconstructed direction.

\subsubsection{\label{sec:scsat}Saturation in scintillator channels}

By restricting our search to scintillator events that saturate, we find that we can achieve a high-purity sample of candidate CR events. Because the energy threshold for detecting radio emission from a CR EAS is much higher than that of the scintillators, as demonstrated in section \ref{sec:sim}, the associated muon energy density may saturate the scintillators in real CR events that contain detectable RF signals. The scintillator gains were not optimized for these highest-energy events of interest.

Saturated scintillator signals deviate from the typical characteristic signal. They quickly rise to or near the maximum ADU voltage and then decay roughly linearly until abruptly returning to the typical exponential decay. We therefore use the zero-to-zero width of the signal as a proxy to flag saturation in a scintillator event instead of amplitude clipping. A characteristic scintillator signal that reaches the maximum ADU voltage exhibits a maximum width of $\sim 260$\,ns, so we classify events as saturated if all four scintillator channels contain a signal wider than this. Note that this may include overlapping signals or atypical unsaturated signals as saturated events, but the rate of these is low at 0.13\% of all COIN-triggered events.

Some properties of the CR EAS would be poorly reconstructed using scintillators that have saturated signals. However, the primary purpose of the scintillators is to provide a trigger for identifying events with RF from CR EAS. We therefore focus the reconstruction of such properties using the RF channels. We still require agreement between the reconstructed directions obtained by the RF and the scintillators (see section \ref{sec:backgrounds}), and saturation affects the uncertainties of the arrival times used in eq. \eqref{eqn:sc_recon} for the scintillator direction reconstruction. The duration of a signal's rising edge is typically $\sim 54.0\,\pm\,9.3$\,ns across all amplitudes $\geq 5$\,ADU, implying a relatively small difference in the 50\% constant-fraction-discriminator threshold that determines the arrival time used in direction reconstruction, which approximately results in a 3$^\circ$ error.

\begin{figure*}[!t]
    \centering
    \includegraphics[width=0.98\linewidth]{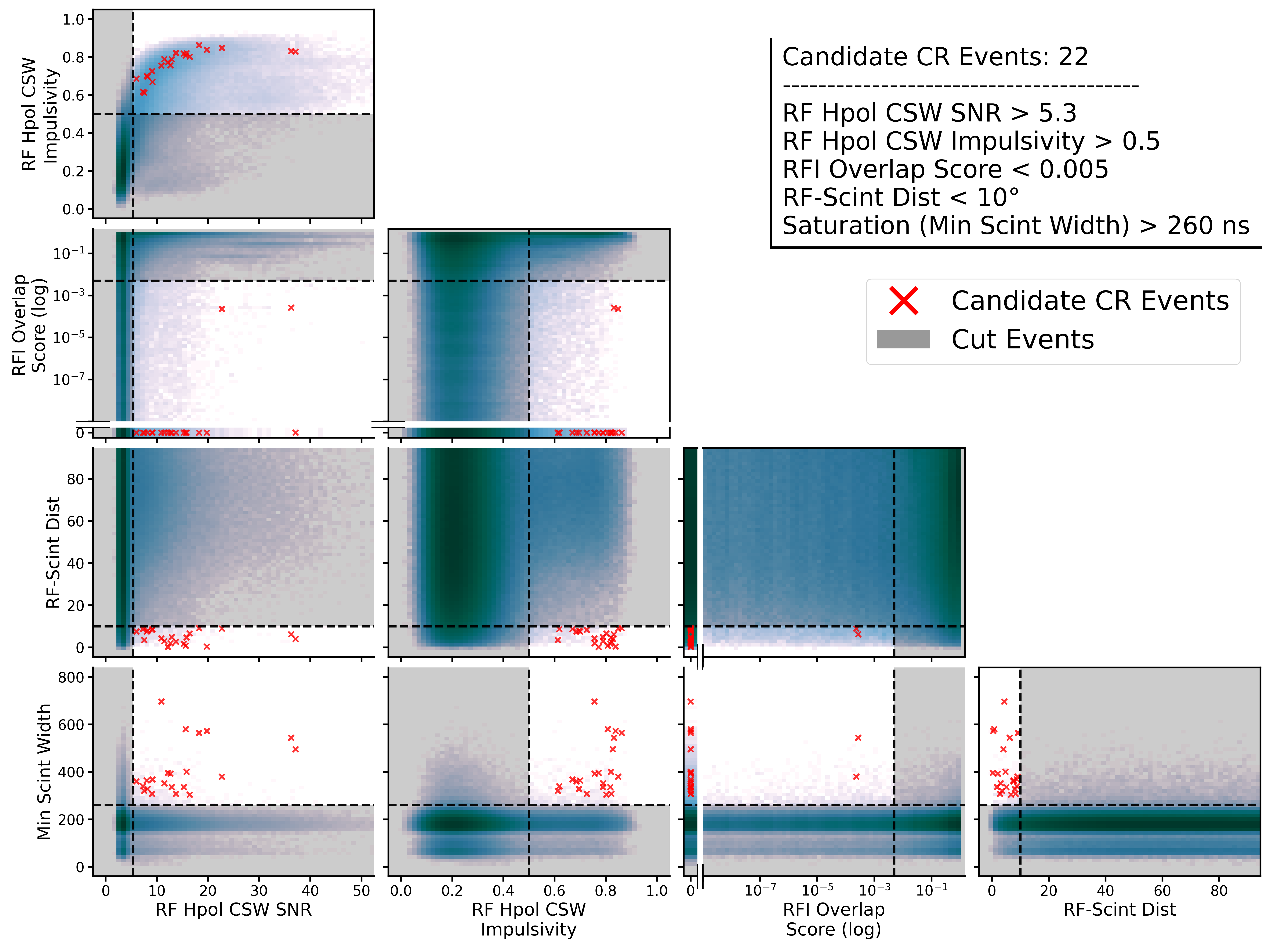}
    \caption{\label{fig:analysisvars}The 2D pairwise distributions of events along the defined analysis variables for the COIN-triggered dataset. The resulting 22 candidate CR events from applying the listed analysis cuts are highlighted in red. The shaded regions of events rejected by the cuts contain the bulk of background events. The angular distance between the reconstructed directions of the radio and scintillator arrays is denoted by ``RF-Scint Dist''. Note that distributions may extend beyond the displayed plot bounds. All histogram counts are log-scaled.}
\end{figure*}

\subsection{\label{sec:backgrounds}Analysis cuts and background estimate}

\begin{figure*}[!t]
    \centering
    \includegraphics[width=0.95\linewidth]{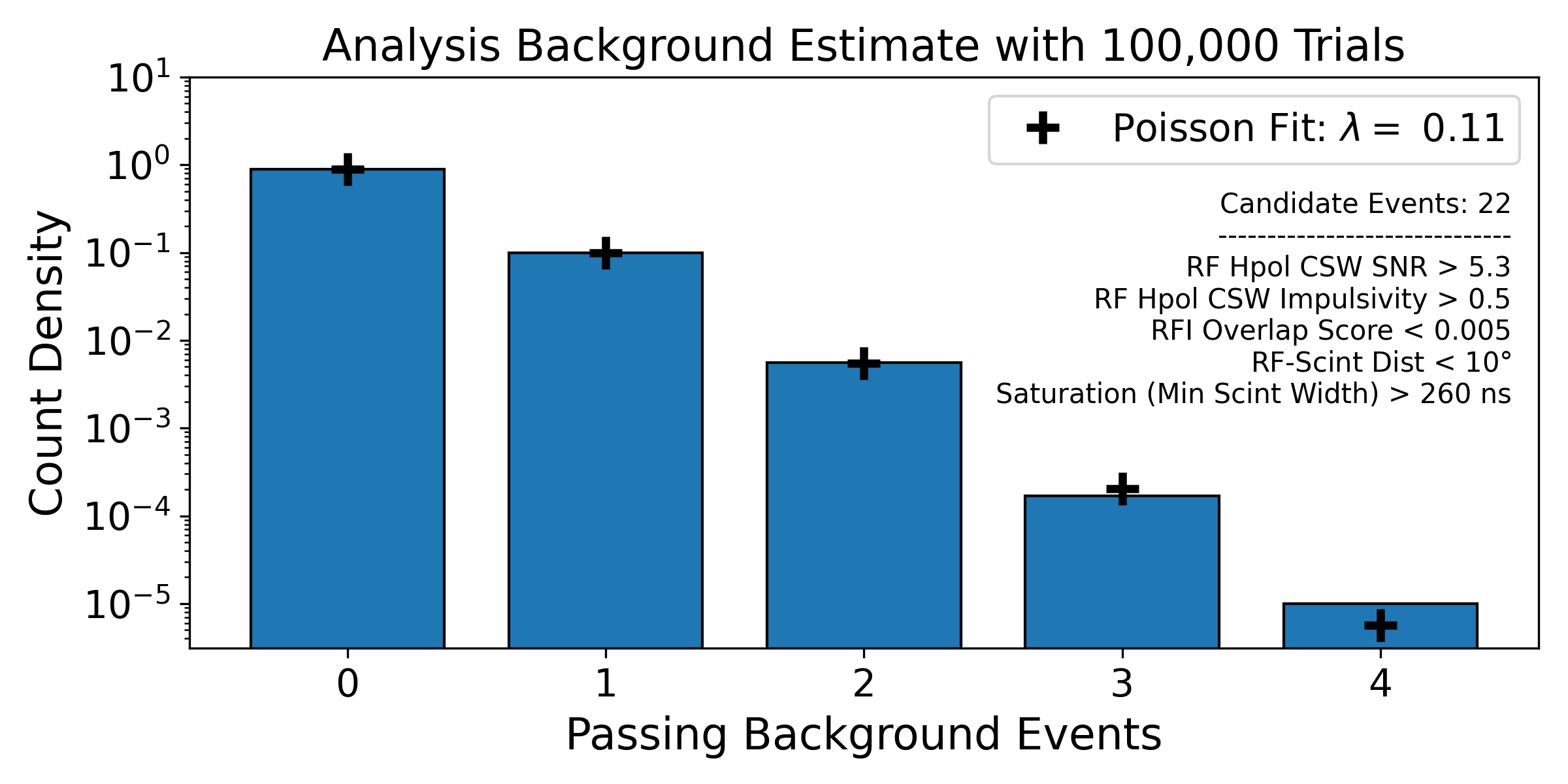} 
    \caption{\label{fig:backgroundestimate}The distribution of synthetic events within 100,000 datasets of randomly paired, uncorrelated radio and scintillator signals which pass our analysis cuts. A Poisson distribution fits it with mean $\lambda=0.11$ events, giving the analysis background estimate for the 22 candidate CR events.}
\end{figure*}

We choose the analysis cuts to optimize the purity of the sample. Figure \ref{fig:analysisvars} shows the distributions of events along the described analysis variables. For example, thermal RF events cluster at impulsivity values below $0.5$ and at Hpol CSW SNR values below $5.3$. The RFI spatial masking selects events with RFI overlap scores below $0.005$. We select events whose reconstructed directions by the radio and scintillator arrays agree within $10^\circ$. The relatively wide 10$^\circ$ agreement on the reconstructed directions is informed by the uncertainties from the RF pointing resolution discussed in section \ref{sec:fieldcal}, the particle plane-wave approximation discussed in section \ref{sec:screcon}, and scintillator saturation discussed in section \ref{sec:scsat}.

We estimate the background of impulsive RF signals unassociated with CRs that are coincident with a scintillator-triggered signal. We do this by applying the analysis cuts to 100,000 synthetic datasets of events with uncorrelated radio and scintillator signals built by randomly pairing RF channels from SW-triggered events with scintillator channels from COIN-triggered events. The number of surviving events in each trial is described by a Poisson distribution, as shown in figure \ref{fig:backgroundestimate}.

\begin{figure*}[t]
    \centering
    \includegraphics[width=0.85\linewidth, valign=c, trim={0 0 0 0},clip]{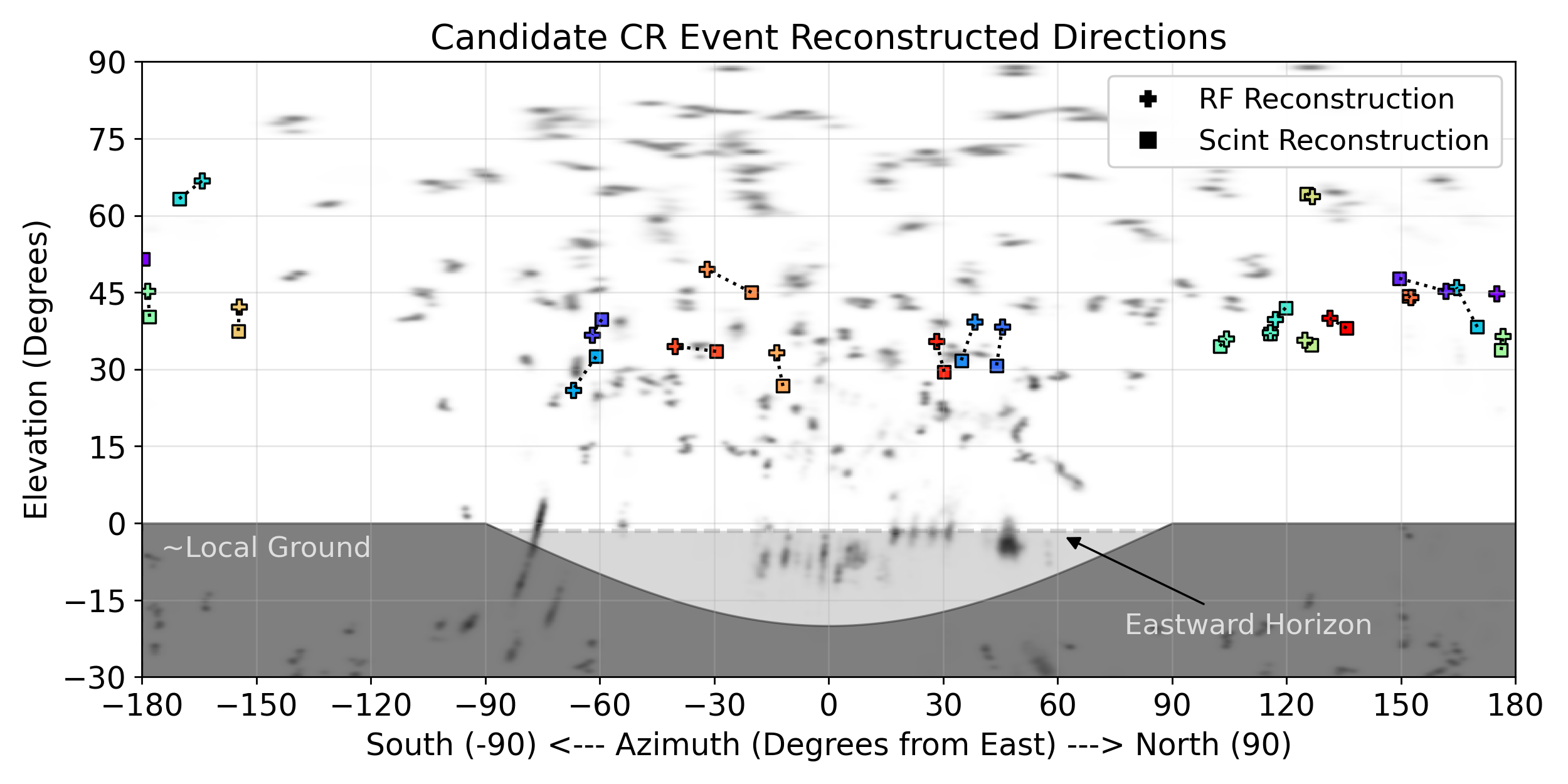}
    \includegraphics[width=0.12\linewidth, valign=c]{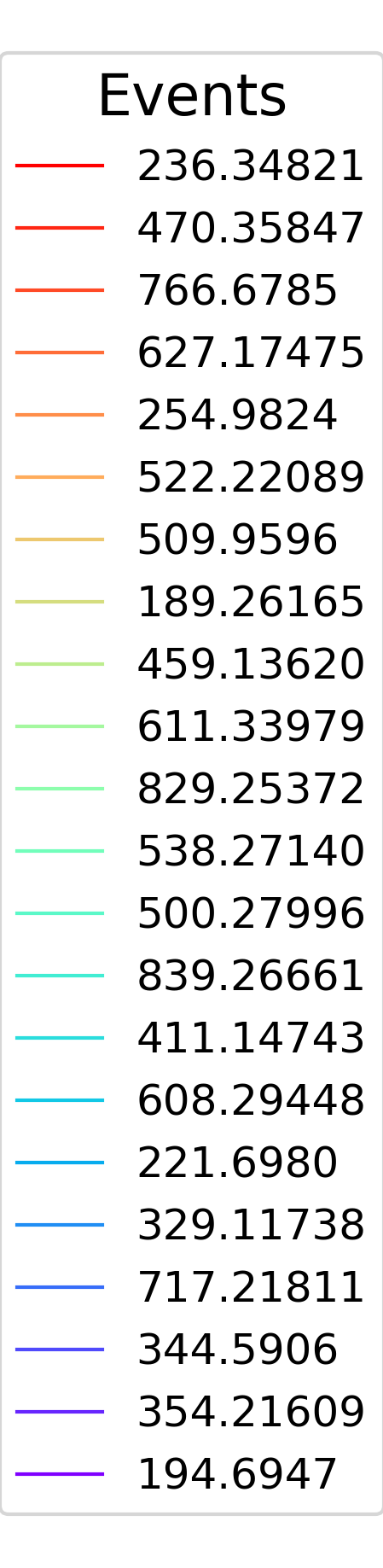}
    \includegraphics[width=0.49\linewidth, valign=c, trim={1.5cm 0 1.2cm 0},clip]{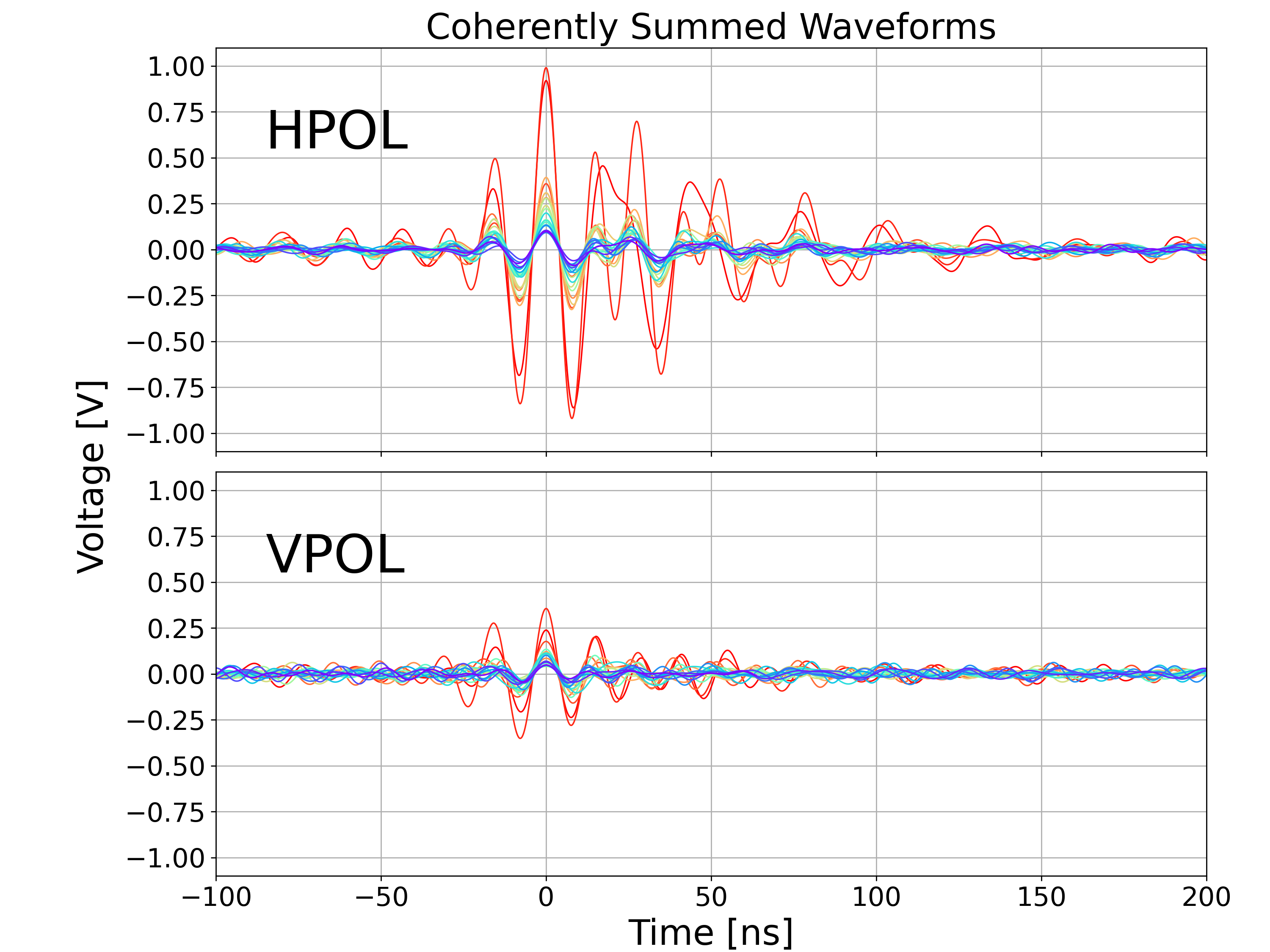}
    \includegraphics[width=0.49\linewidth, valign=c, trim={1.5cm 0 1.2cm 0},clip]{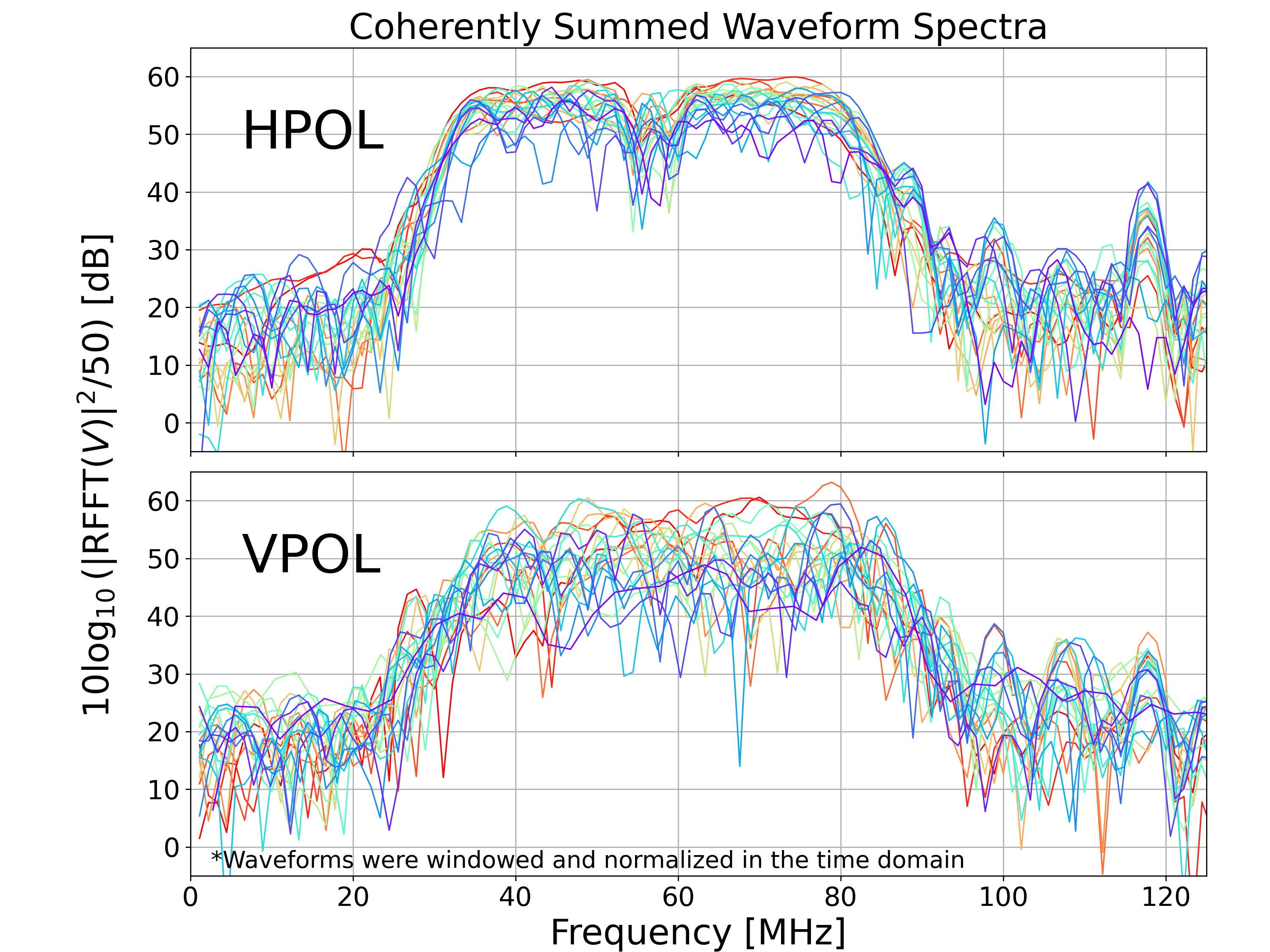}
    \caption{\label{fig:crevents}Top: Reconstructed arrival directions of the 22 candidate CR events overlaid on the RFI distributions defined in section \ref{sec:rfbg}. Dotted lines connect the RF and scintillator reconstructed directions for the same event (except for event 194.6947 due to wrap-around). Bottom: The CSWs (peaks aligned at t=0\,ns for visualization) and their frequency spectra for the 22 candidate CR events. Events are ordered by their Hpol CSW SNR.}
\end{figure*}

We identify 22 candidate CR events on an estimated background of $0.11$ events. The resulting fit on the background estimate has an associated $\chi_\text{red}^2 = 0.23$ and $p\text{-value} = 0.79$ with two degrees of freedom.

We note that saturation in the scintillator channels results in higher purity of the identified CR sample. Without saturation, we identify 123 events on an estimated background of $86.44$ events. Including the scintillator saturation cut reduces the number of candidate CR events to 18\%, but reduces the background estimate further to 0.1\%.

The resulting 22 candidate CR events are shown in figure \ref{fig:analysisvars} among the full COIN-triggered dataset distributions. The sky map of reconstructed directions in both the RF and scintillators for the candidate CR events is shown at the top of figure \ref{fig:crevents}, which also shows an overlay of each event's CSW in time and frequency spectrum.

Note that previous analyses restricted CR searches to azimuths between $-70^\circ$ and $+70^\circ$, constrained by the angular coverage of the available E-field simulation libraries. The 2021 instrument likewise only used RF-trigger beams within this range \cite{danthesis-2022PhDT19S, andrewthesis-Zeolla:2026pan}. Using the scintillators as an independent trigger, we are able to identify radio-detectable CR EAS emission at azimuths beyond $\pm70^\circ$ of the instrument.

\subsubsection{\label{sec:airplane}Coincidence with airplanes}

\begin{figure*}[!t]
    \centering
    \includegraphics[width=0.9\linewidth]{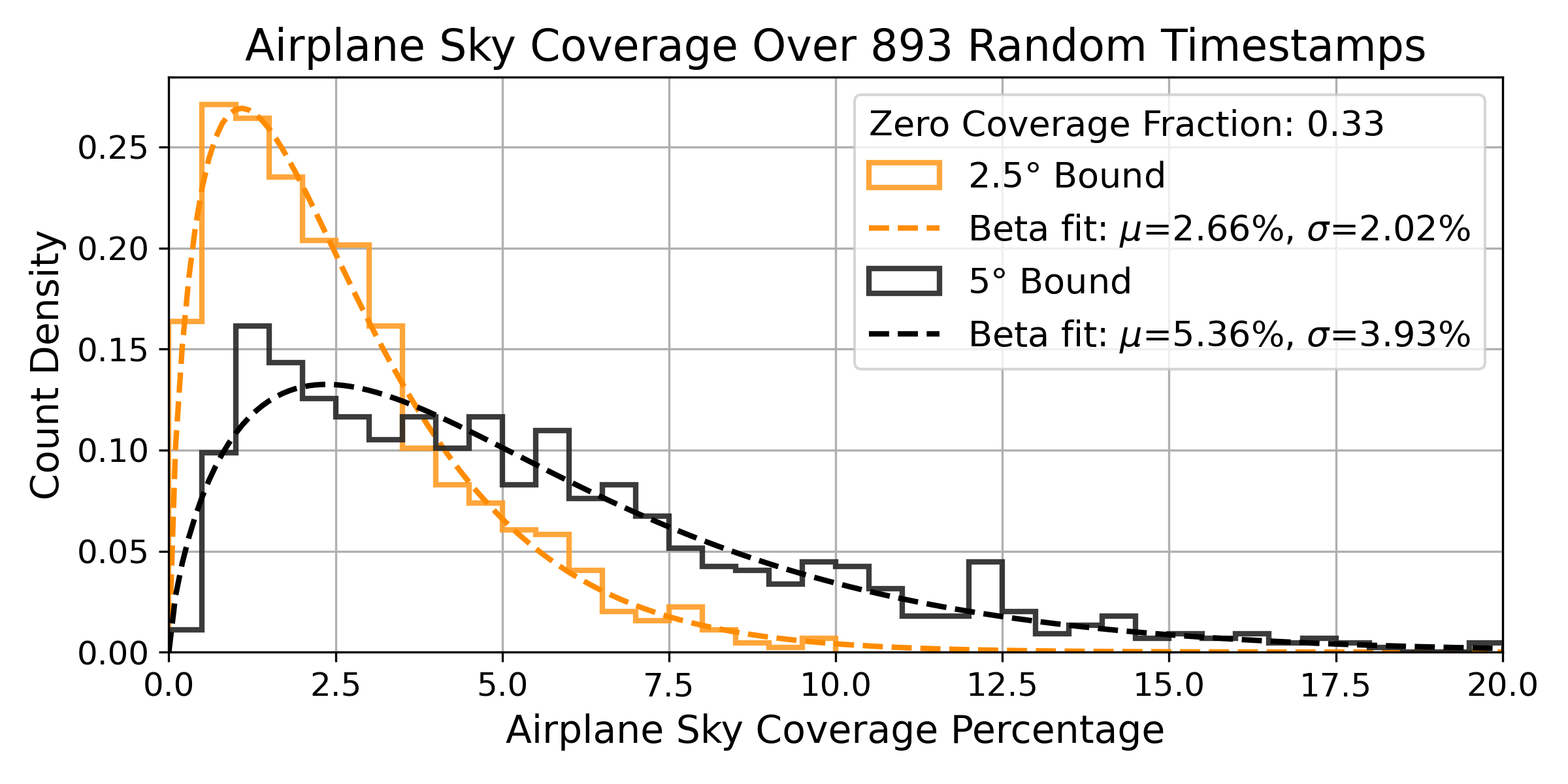}
    \includegraphics[width=0.9\linewidth]{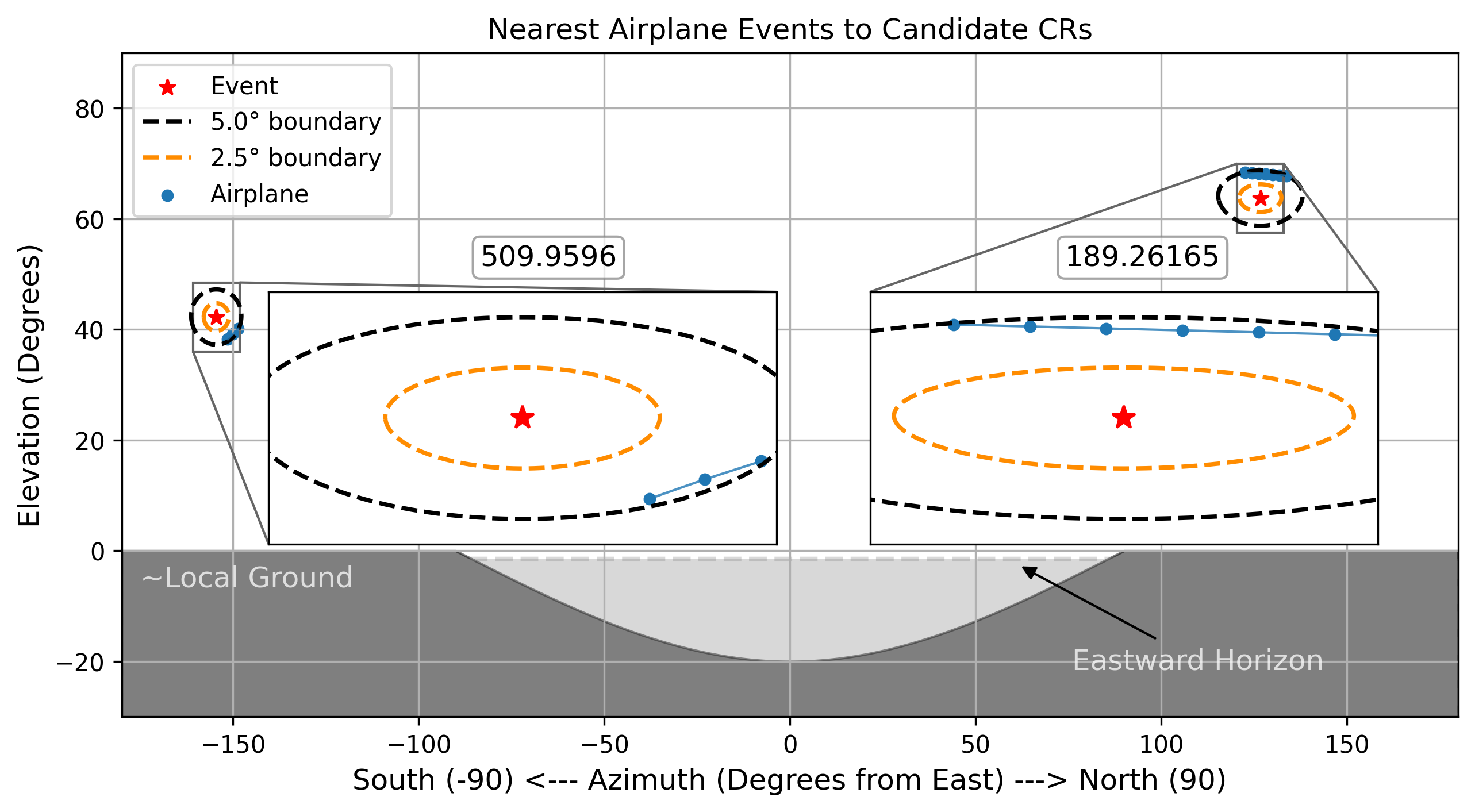}
    \caption{\label{fig:airplanes}Top: The normalized distributions of nonzero sky coverage by airplanes using 2.5$^\circ$ and 5$^\circ$ bounds on airplane tracks identified in randomly sampled timestamps. The fraction of sampled timestamps where there is zero coverage is 0.33. Bottom: Two candidate CR events plotted on the sky in proximity to airplane tracks with the 5$^\circ$ pointing boundary, but not with the 2.5$^\circ$ boundary.}
\end{figure*}

As mentioned in section \ref{sec:rfbg}, we cross-check directions and timestamps of candidate CR events against nearby airplanes since the RFI clustering method used for RFI rejection does not capture track-like signals exhibited by airplanes. For each candidate CR event, we check, using data from OpenSky Network \cite{opensky-6846743}, for the presence of any airplane within 50\,km of the array, within a $\pm$1 minute time window of the event timestamp, and whose track passes within 2.5$^\circ$ or 5$^\circ$ of the event reconstructed direction. 

The 2.5$^\circ$ and 5$^\circ$ bounds represent estimates of our pointing resolution for airplanes, using the calibrated pointing distribution shown in figure \ref{fig:cal}, with the 5$^\circ$ bound being the most conservative. As shown at the bottom of figure \ref{fig:airplanes}, we find two events that point near airplanes within the $\pm1$ minute window and 5$^\circ$ bound, but none within the 2.5$^\circ$ bound. 

We estimate how likely it is that two CR events coincide with airplanes by characterizing the sky coverage by airplanes at any given time, and comparing it to the expected distribution of CRs (see section \ref{sec:sim}). We calculate the sky coverage by randomly sampling timestamps in the data, identifying present airplane tracks within $\pm$ 1 minute of the timestamp, upsampling those tracks, and performing a Monte Carlo that samples random points in the expected spatial CR distribution. We define the airplane sky coverage as the fraction of sampled CR positions that lie within 2.5$^\circ$ or 5$^\circ$ of any airplane track point at a given timestamp.

The distributions of resulting nonzero sky coverages across 893 random timestamps for a 2.5$^\circ$ or 5$^\circ$ bound are shown at the top of figure \ref{fig:airplanes}. We fit these with Beta distributions. No airplane tracks are identified for 33.04\% of the sampled timestamps. The effective coverage is the mean coverage from the distribution in figure \ref{fig:airplanes} multiplied by the fraction of nonzero coverage samples. We estimate the expected accidental airplane-coincidence count by multiplying the number of candidate CR events by the effective coverage. Thus, for a 2.5$^\circ$ angular boundary, the mean airplane coincidence count is 0.39 $\pm$ 0.30 and for 5$^\circ$, this is 0.79 $\pm$ 0.58. For the two events identified within 5$^\circ$ of airplanes, this corresponds to a significance of 0.9$\sigma$. Considering that 5$^\circ$ is the conservative bound, and that these two events exhibit characteristics of real CR events, we do not classify the two events as airplanes and keep them both as candidate CR events during event characterization in section \ref{sec:cr}.



\section{\label{sec:sim}Simulated hybrid-instrument performance}

We simulate the qualitative features of the hybrid-instrument performance, targeting CR event observations that would trigger the scintillator detectors and be detectable by the radio array. The simulations utilize the combination of measured and modeled system responses for the scintillator and radio arrays. Analysis cuts are applied to hybrid events, or events that would trigger both arrays independently, similarly to those done in section \ref{sec:analysis}, in order to compare the resulting event rate and spatial distribution with those obtained from the 22 candidate CR events.

\begin{figure*}[!t]
    \centering
    \includegraphics[width=0.95\linewidth]{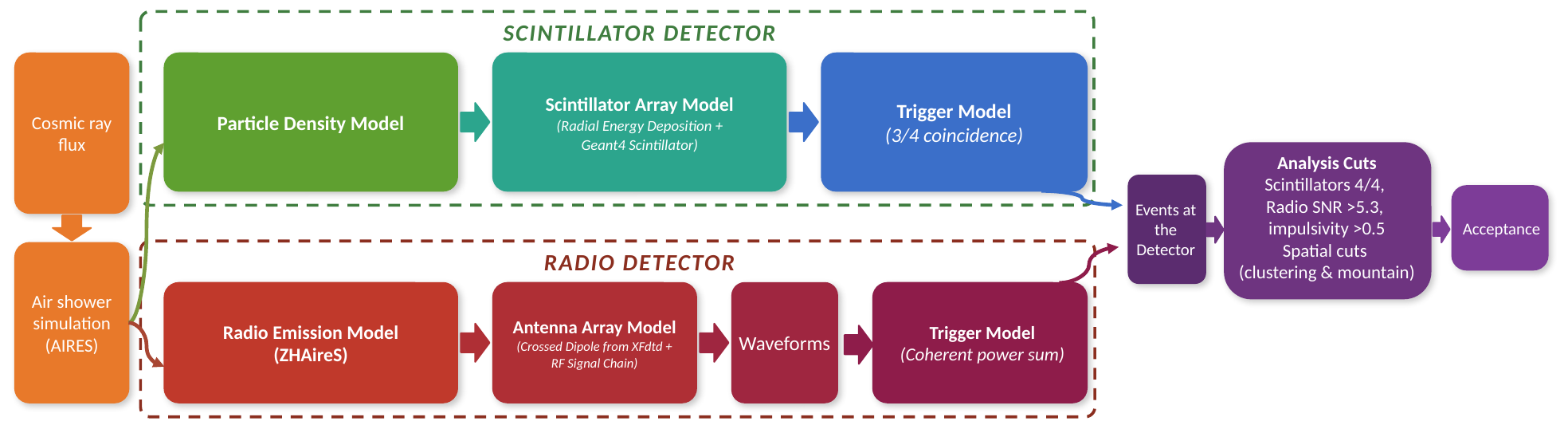}
    \caption{\label{fig:flowchart}Flowchart of the CRANBERRY simulation, highlighting the independent radio and scintillator trigger branches for the 2023 BEACON prototype, and selection for simulated hybrid events. Hybrid events are chosen from CR showers that trigger the scintillator array and are then selected if they also produce an associated radio signal that triggers the radio array. Analysis cuts meant to mimic the event selection in the analysis described in section \ref{sec:analysis} are also applied to finalize the simulated hybrid event set. More details on the simulation setup are included in the text.}
\end{figure*}

We use the CRANBERRY simulation framework~\cite{Zeolla:2023, andrewthesis-Zeolla:2026pan} to model the BEACON prototype's instrument responses to a CR EAS. CRANBERRY is a Monte Carlo simulation package initially developed with only a radio array model to determine the acceptance of the BEACON prototype to CRs, which we extend to include the scintillator array for this work. A flowchart of the CRANBERRY pipeline with the scintillator extension is shown in figure \ref{fig:flowchart}.

For the radio array acceptance, CRANBERRY first generates CR EAS events with randomized incident directions within a fixed radius around the array. The radio emission of each CR is then interpolated using a library of CR EAS simulations generated using \texttt{ZHAireS} \cite{zhaires_Alvarez_Mu_iz_2012} and forwarded through the BEACON signal chain.

To extend CRANBERRY to calculate the scintillator array acceptance, we include the lateral distribution of particles at the ground recorded in a CR EAS simulated by \texttt{ZHAireS}. Each particle, tracking its direction, energy, and weight, is fed into a Geant4 \cite{geant4} simulation modeling the interaction through a scintillator detector model, following the description in section \ref{sec:scint}, to determine its deposited energy into a scintillator. Using the energy deposited by each particle in the EAS and their positions within the lateral distribution, we build a radially binned map of energy deposition density centered around the shower axis. Through this procedure, we build a database of simulated CR EAS events at a primary energy of $10^{17}$ eV over a grid of elevation $\theta\in[0,90]^\circ$ and azimuth $\phi\in[-180,180]^\circ$ with $5^\circ$ angular increments. At energies above the radio acceptance threshold, $10^{16} \mathrm{eV}\leq E_\mathrm{CR} \leq 10^{20}\mathrm{eV}$, where hybrid-detected events are expected, the scintillator energy deposition map scales linearly with $E_\mathrm{CR}$.

\begin{figure*}[!t]
    \centering
    \includegraphics[width=0.46\linewidth]{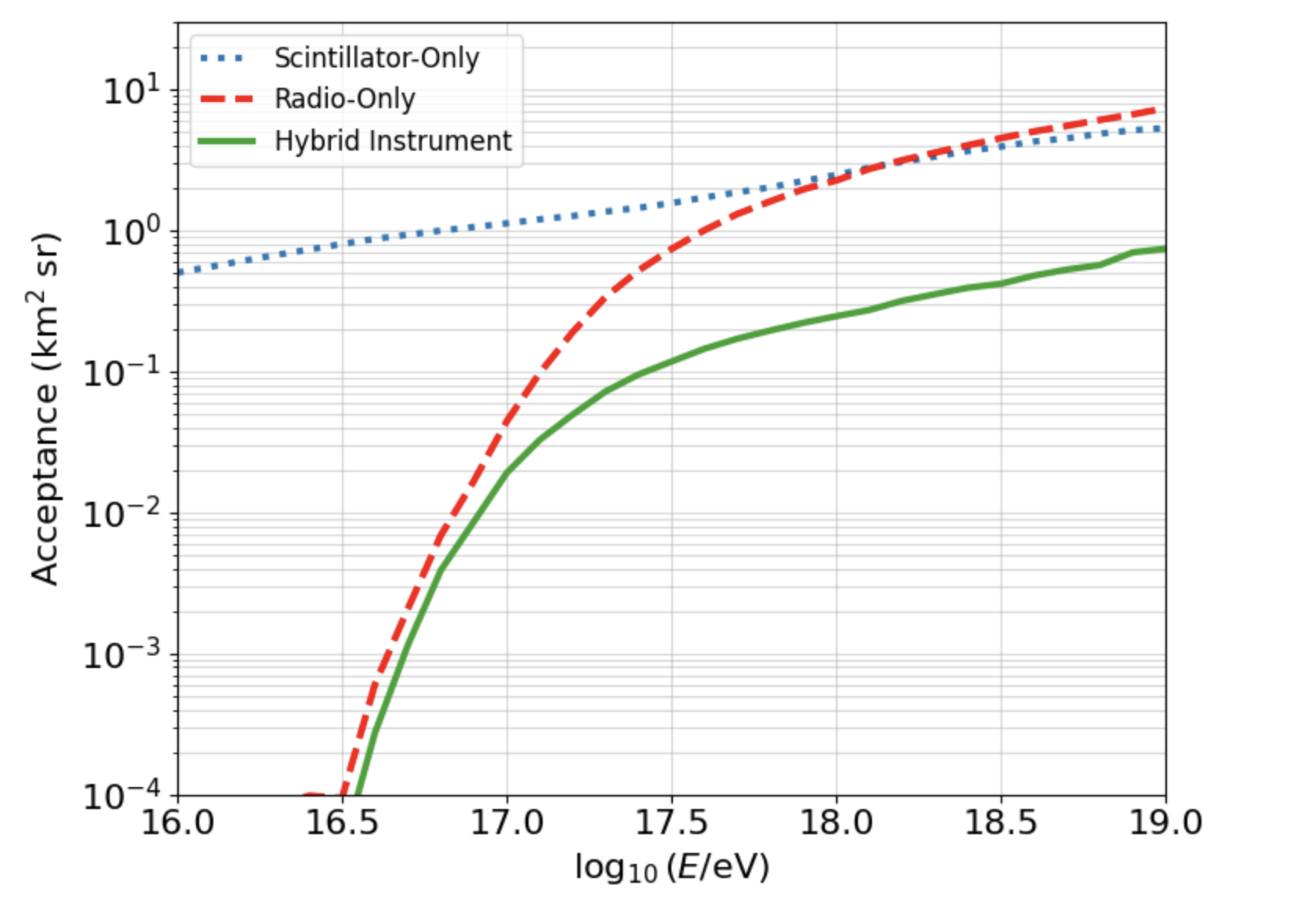}
    \includegraphics[width=0.50\linewidth]{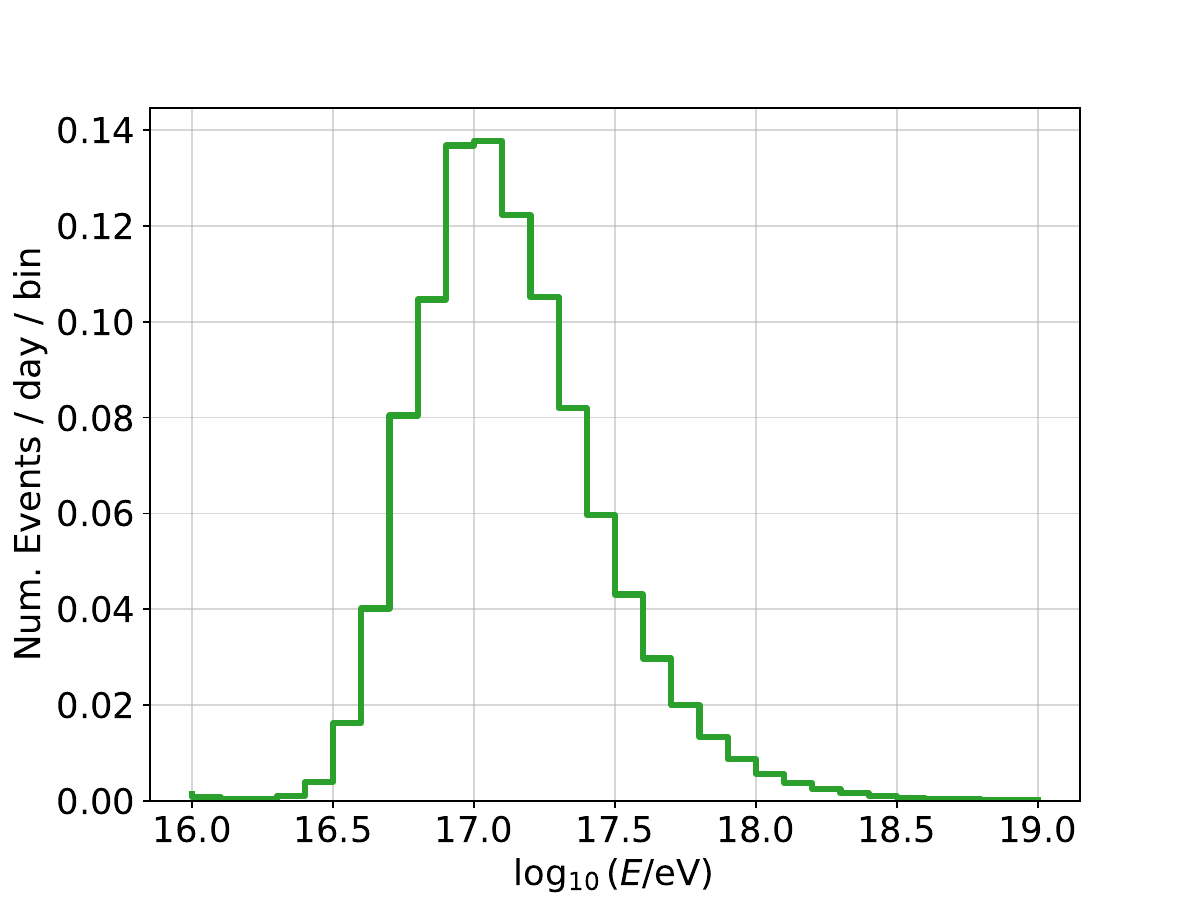}
    \caption{\label{fig:acceptance_eventrates}Left: The resulting simulated acceptances from CRANBERRY, comparing all events that trigger the scintillator array (i.e. only the scintillator branch), all events that trigger the radio array (i.e. only the radio branch), and all events that trigger both arrays and pass analysis cuts (hybrid events). Right: The event rate expected for hybrid events, resulting from the hybrid instrument acceptance and expected CR flux. The total number of detected events expected per day is $\sim 1$.}
\end{figure*}

For each simulated CR event sampled by CRANBERRY, we identify the database entry with the closest matching geometry and scale its corresponding map with $E_\mathrm{CR}$. The energy deposition into each scintillator in the array is determined by extracting the local density at the position of each detector and multiplying by the scintillator's effective surface area with respect to the shower axis, 
\begin{equation} \label{eq:effective_scint_area}
    A_\mathrm{eff,sc}=1.5\,[\mathrm{m^2}]\,\cdot\,\big(\sin\theta\cos15^\circ - \cos\theta\sin\phi\sin15^\circ\big)~,
\end{equation}
where $\theta$ and $\phi$ are the event elevation and azimuth angles, respectively, and $15^\circ$ corresponds to the scintillator tilt. The equivalent signal in each scintillator is calculated using eq. \eqref{eq:scint_energy_conversion}.

The simulated instrument mimics a trigger by passing simulated CR events that, for the radio array, have coherently summed waveform SNR greater than 5.3, and, for the scintillator array, have at least 3 out of 4 scintillators see a deposited energy signal above $0.75 \:\mathrm{ADU-}\mu\mathrm{s}$, which approximately corresponds to the $5\:\mathrm{ADU}$ peak used in the real scintillator coincidence trigger on the 2023 BEACON prototype.

We simulate passing hybrid events in the 2023 BEACON prototype using CRANBERRY by forwarding each simulated CR event through the radio and scintillator branches separately, and identifying which events would trigger both arrays. 500,000 CR events were simulated per energy bin. To each event, we apply analysis cuts similar to those in section \ref{sec:analysis}. For the radio branch, we require that events have a CSW SNR $> 5.3$ and impulsivity $> 0.5$. For the scintillator branch, we require that all four detectors have an integrated signal $>16\:\mathrm{ADU-}\mu\mathrm{s}$, approximately equivalent to the minimum pulse width requirement of the scintillator saturation cut. A cut on the arrival direction of the passing events is applied using the RFI overlap scores defined in section \ref{sec:rfbg}. A cut is also applied to select events which do not point towards the direction of the local mountain ground at the prototype site, determined by the condition,
\begin{equation}
    \theta > 30^{\circ}\cdot|\cos\phi|~,
\end{equation}
on events with direction beyond $\phi \in(-90,90)^\circ $, treating the mountain as a $30^{\circ}$ ground plane.

The left panel of figure \ref{fig:acceptance_eventrates} presents the simulated acceptances of the 2023 BEACON prototype instrument as a function of CR primary energy, and compares the hybrid event acceptance to that of the detector arrays individually. The right panel of the figure shows the expected event rates per energy bin for the hybrid events, given the cosmic ray flux measured by the Pierre Auger Observatory \cite{Fenu:2017OO}. When integrated over its energy range, the total event rate for the hybrid instrument is $\sim 1 \: \: \mathrm{event/day}$, compared to the $\sim 0.3 \: \: \mathrm{events/day}$ rate found in this analysis after all analysis cuts were applied.  Although we applied cuts to simulate the effects of our analysis on the simulated data rate, the cuts were not identical to those used in analysis, importantly including the saturation cut, which is difficult to simulate. Additionally, the predicted event rate from simulation is sensitive to the exact antenna model used, and assumptions about the continuum RF noise levels, which lead to uncertainty on the simulated rate. We take these rates to be in good agreement between the observed and simulated event rates. 

\begin{figure*}[!t]
    \centering
    \includegraphics[width=0.32\linewidth]{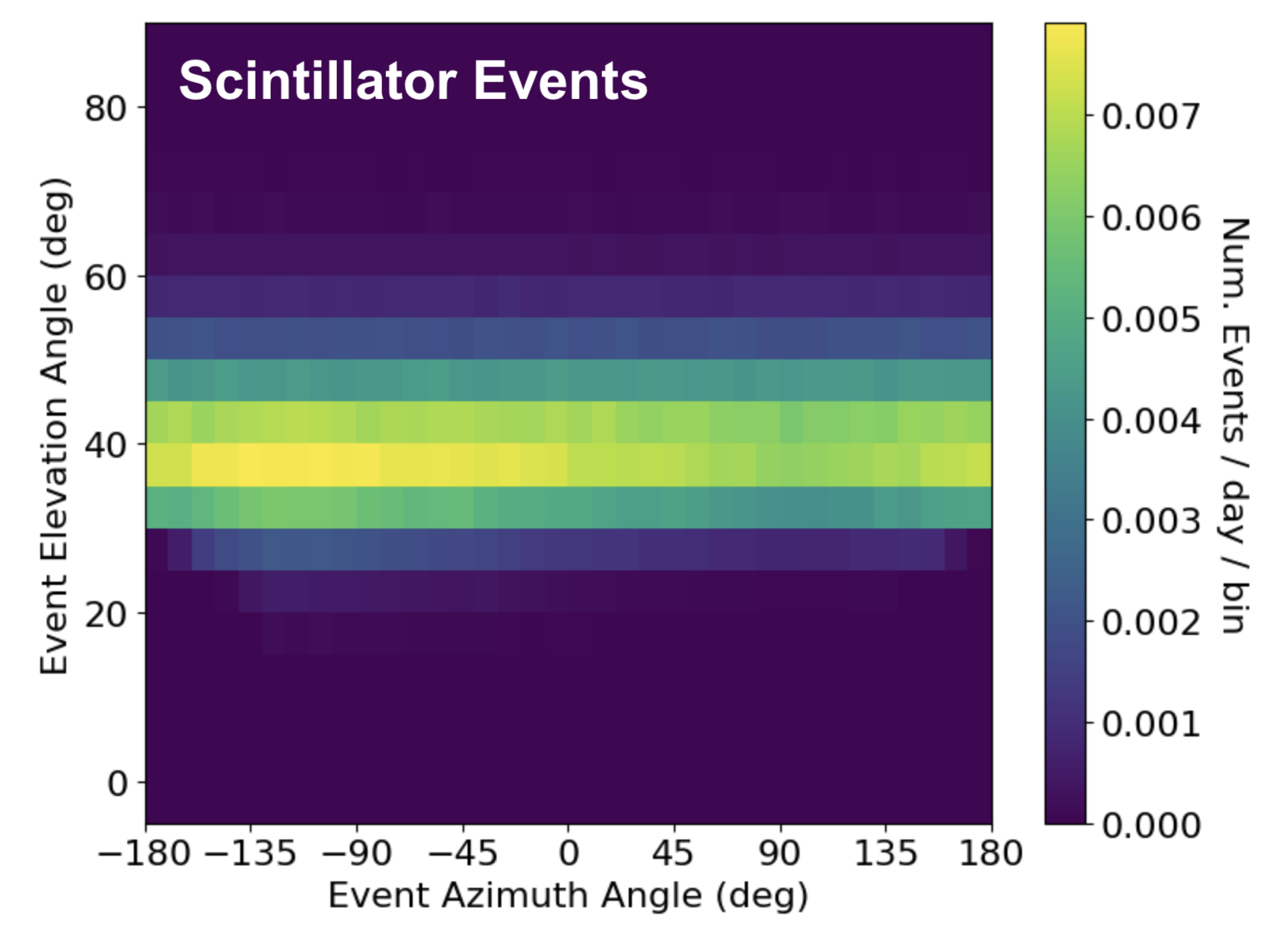}
    \includegraphics[width=0.32\linewidth]{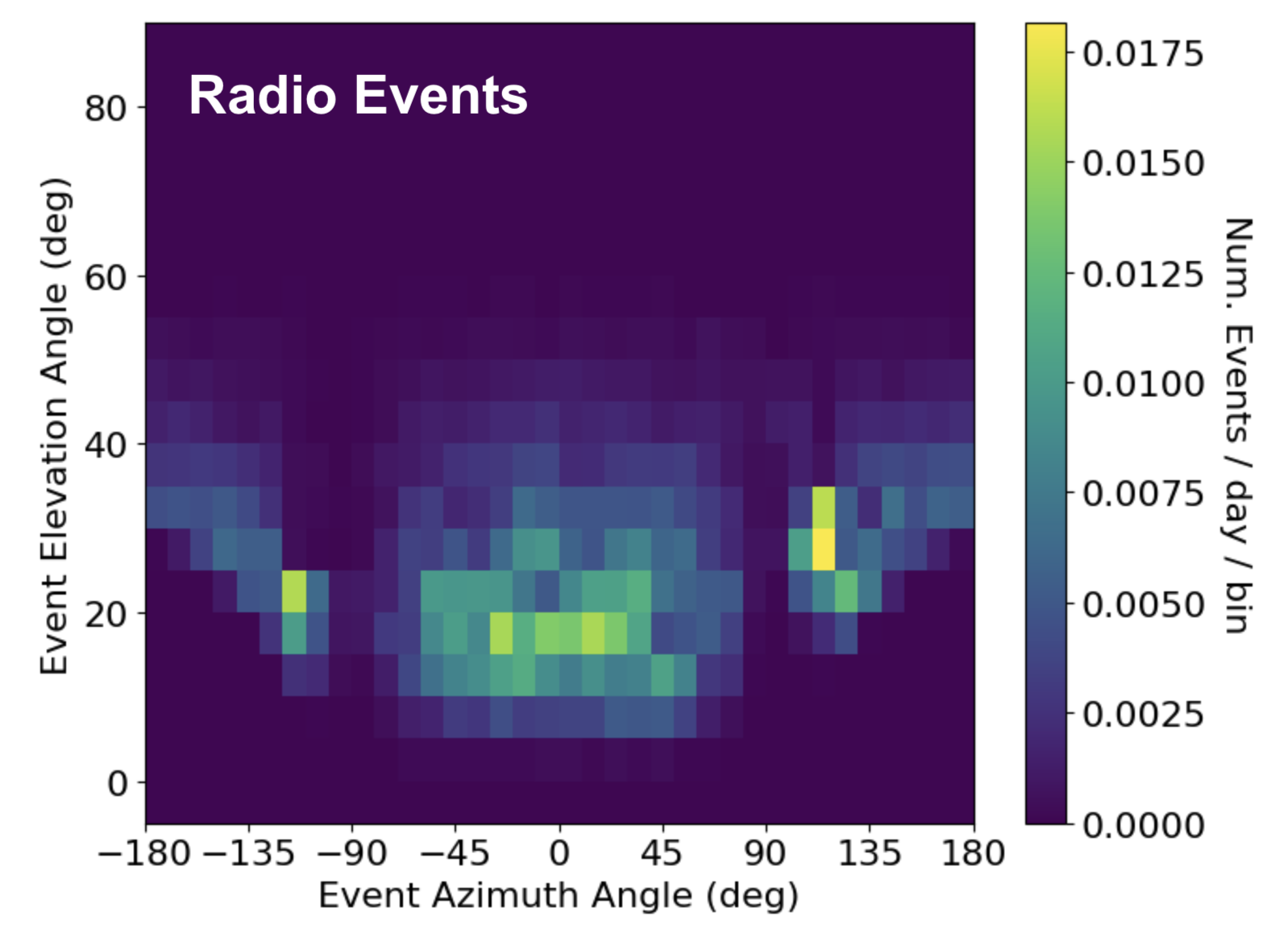}
    \includegraphics[width=0.32\linewidth]{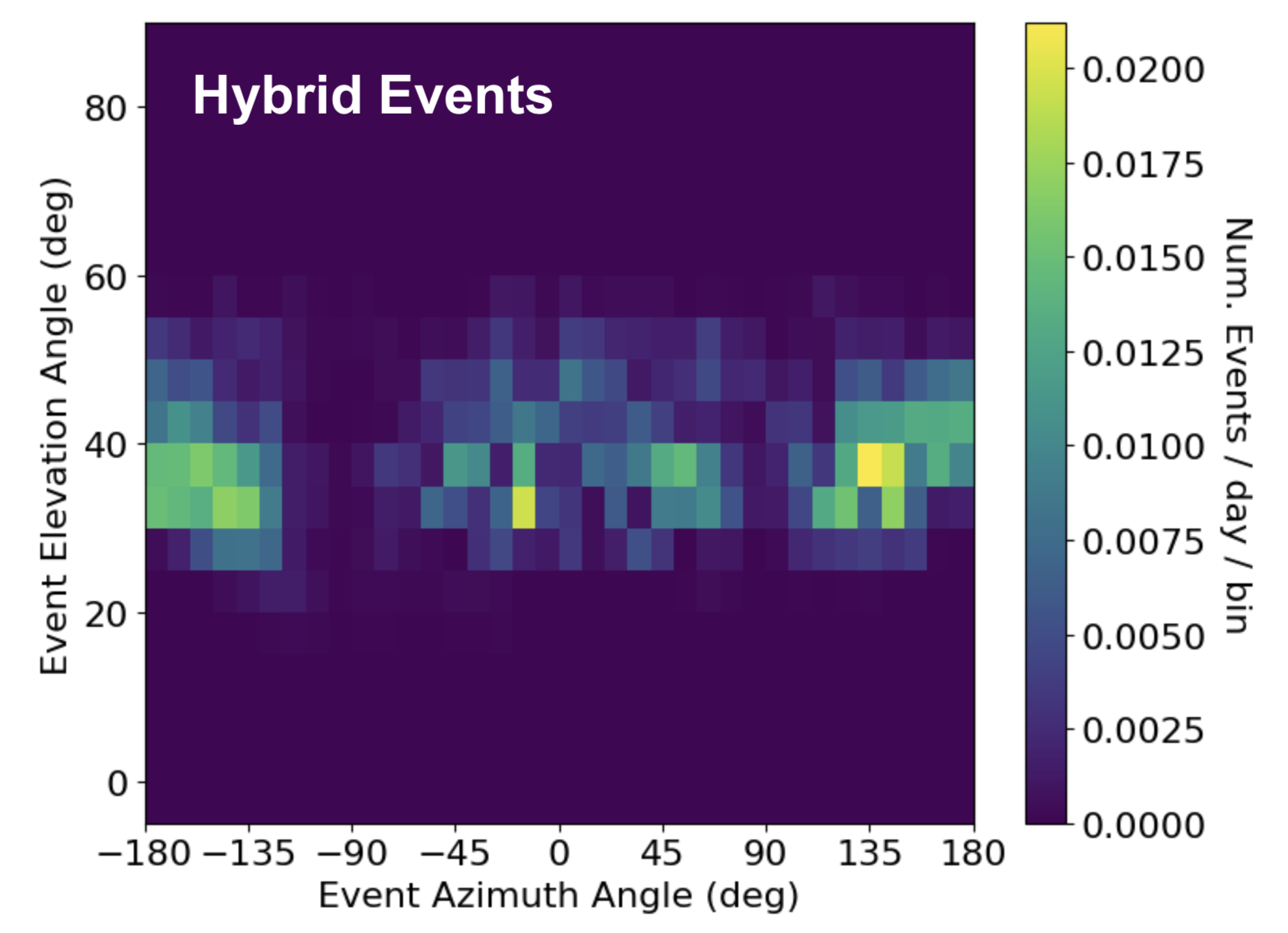}
    \caption{\label{fig:simskymap} The spatial distributions across energies $10^{16} \mathrm{eV}\leq E_\mathrm{CR} \leq 10^{19}\mathrm{eV}$ of simulated events that trigger the scintillator array (left), that trigger the radio array (center), and that trigger both arrays and pass analysis cuts (right). At these energies, the maximum of the shower development would occur below the detector for near-vertical showers, which result in the upper elevation bounds. At more inclined angles, the shower occurs far away from the detector, setting the lower elevation bound on muons for the scintillator events, but not the far-traveling radio signals. The presence of the mountain is also included, resulting in the apparent cutoff of events at azimuths beyond $\pm90^\circ$.}
\end{figure*}

Figure \ref{fig:simskymap} displays the spatial distributions on the sky of simulated CR events that trigger the radio and scintillator detectors within the CRANBERRY simulation framework, comparing events passing the individual detectors and hybrid events passing both detectors with analysis cuts applied. On their own, the scintillator array is expected to trigger most efficiently on higher elevation showers while the radio instrument is most efficient for more inclined events. The sky map shows that as a consequence of combining these two different sensitivity regions, the hybrid instrument is sensitive to the band, $\theta \simeq 40^{\circ}\pm10^{\circ}$. A distinctive azimuthal structure is seen with the highest sensitivity near $\phi=0^{\circ},180^{\circ}$, which is a result of the forward-beamed sensitivity of the radio dipole antennas.



\section{\label{sec:cr}Event characterization}

The BEACON prototype's dense and relatively small 50\,m baseline array geometry prevents us from reconstructing the EAS lateral distribution function (LDF) as done by larger CR radio arrays to extract parameters that estimate the CR primary energy, $X_{\text{max}}$, and charge-excess contribution to polarization~\cite{aera-PhysRevD.93.122005, auger_PhysRevD.89.052002, lofarldf-Nelles_2015, lopesldf-Palmieri_2013, ovrolwa-MONROE2020163086, radioGood-HUEGE20161, radioGood-Buitink_2014}. We therefore reconstruct only the local E-field and polarization for the 22 candidate CR events to verify their consistency with expected CR characteristics.

\subsection{\label{sec:efield}Reconstructed electric field}

We reconstruct the E-field seen by each antenna by deconvolving the complex antenna and RF signal chain responses, detailed in section \ref{sec:antennas}, from the received Hpol and Vpol voltage waveforms. Neglecting added receiver noise, the received voltages, $V_\text{rec}$, are related to an incident E-field, $\mathbf{E}_\text{inc}$, by the frequency-domain matrix equation \cite{farr1998timedomain, balanis2016antenna},
\begin{equation}
    \begin{bmatrix}
        V^\text{H}_\text{rec} \\
        V^\text{V}_\text{rec}
    \end{bmatrix}
    = 
    \sqrt{\frac{Z_c}{Z_0}}
    \begin{bmatrix}
        \eta^\text{H}\,\mathbf{H}_N^\text{H} \\
        \eta^\text{V}\,\mathbf{H}_N^\text{V}
    \end{bmatrix}
    \cdot
    \mathbf{E}_\text{inc}
    ~,
    \label{eqn:efield}
\end{equation}
where the full response matrix is composed of the measured signal chain response, $\eta$, and the simulated complex antenna response, $\mathbf{H}_N$, from eq. \eqref{eqn:antresp}. The reconstruction computes a regularized inversion of the response matrix in the 30 -- 80 MHz band, preventing numerical artifacts where the response is near zero (e.g. out-of-band).

From each channel's signal chain response, measured at 620 kHz frequency resolution, we remove the long-cable group delay to reduce oscillations in the complex values. We choose the appropriate antenna response from section \ref{sec:antennas} according to the antenna location and the reconstructed direction of the candidate CR event, and Fourier-resample it to a finer frequency grid than the 5 MHz spacing provided by the simulation. We interpolate all responses onto the data frequency grid using linear interpolation of the real and imaginary components.

Among the 22 candidate CR events, the mean reconstructed E-field peak amplitude measured in the antennas is $\sim 300\,\mu$V/m, with some amplitudes as high as $\sim 3000\,\mu$V/m.

\subsection{\label{sec:pol}Reconstructed Polarization}

\begin{figure*}[!t]
    \centering
    \includegraphics[width=0.8\linewidth]{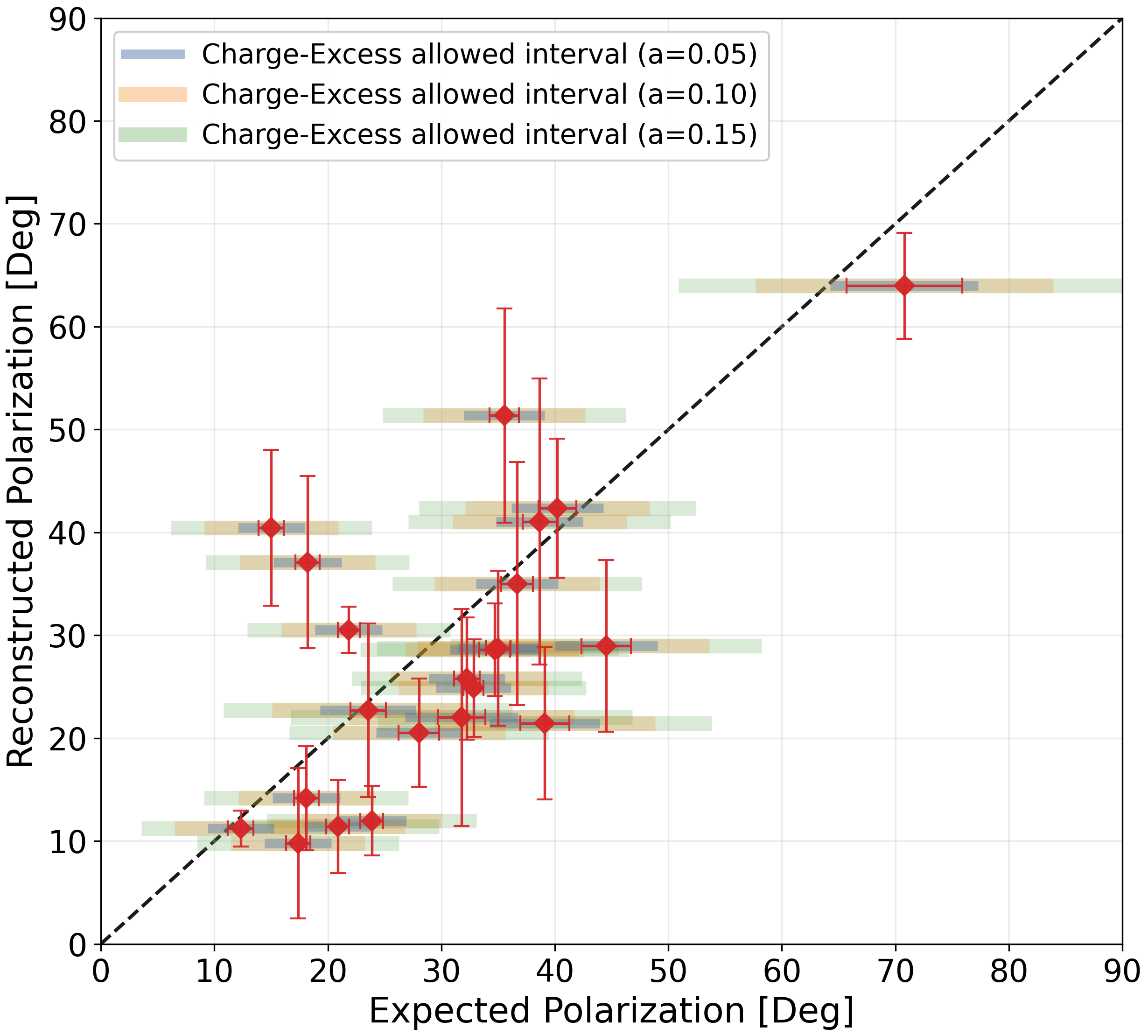}
    \caption{The unsigned principal polarization angles reconstructed for the 22 candidate CR events. Associated uncertainties are discussed in detail in section \ref{sec:pol_unc}. The horizontal bars indicate intervals within which a charge-excess contribution on the E-field may cause the expected polarization angle to deviate from the pure geomagnetic value, with shower-to-magnetic-field $\sin(\alpha)$-corrected amplitude factors of $a = 0.05, 0.10, 0.15$, defined by eq. \eqref{eqn:a_ask}.}
    \label{fig:polarization}
\end{figure*}

For a reconstructed arrival direction, we transform the E-field components computed for each event, using eq. \eqref{eqn:efield}, into their azimuth and elevation components. The resulting calculated polarization angles are therefore relative to the azimuth basis vector of the spherical coordinate system.

We assume that the reconstructed arrival direction approximates the shower axis and that the polarization angle does not change significantly across the array. The Hpol channel of the antenna has varying sensitivity to the azimuth component, but this is accounted for in the antenna response.

Due to the dispersion of the signal from band limiting, the determination of the sign of the reconstructed E-field is unreliable on an antenna-by-antenna basis. With just six antennas, we consider the unsigned principal polarization angle, ranging from 0$^\circ$ to 90$^\circ$, using the absolute values of the amplitude ratios of the E-field components.

We derive the unsigned principal polarization angles using the Stokes parameters computed from the analytic waveforms of the two E-field components~\cite{stokes,auger_PhysRevD.89.052002,Chiche_2022}, 
\begin{equation}
    \phi_p = 0.5\,|\mathrm{arctan2}\left(U,Q\right)|~,
\end{equation}
where
\begin{align}
    Q &= \langle |E_\phi |^2\rangle - \langle |E_\theta|^2\rangle \\
    U &= 2~\text{Re}\{\langle E_\phi E^*_\theta\rangle\}
\end{align}
are linear-polarization Stokes parameters.

The degree of linear polarization, determined by the phase offset of the E-field components, can, in principle, be used to distinguish anthropogenic RFI from real CR events, but given the uncertainties in time alignment between polarization channels, we instead enforce a phase offset of zero by aligning the E-field component peaks with a restricted cross-correlation. This is done by restricting the contributing portion of the E-field to the full width at half maximum (FWHM) of the peak in the Stokes intensity parameter,
\begin{equation}
    I = |E_\theta(t)|^2+|E_\phi(t)|^2 ~.
\end{equation}
In low-SNR events, additional noise peaks may occur within 50 ns of the primary peak, effectively extending the window. The primary peak has a typical FWHM of approximately 20 ns, but the window may reach 80 ns when multiple peaks are included.

We compare the resulting polarization angles with the expectation from pure geomagnetic emission, the dominant radio emission mechanism for an EAS \cite{emission-Schr_der_2017}. We compute this expected angle using the reconstructed shower velocity vector, $\mathbf{v}$, and Earth's magnetic field, $\mathbf{B}$, at the site's location based on the World Magnetic Model data \cite{Chulliat2025WMM} retrieved through \href{https://pypi.org/project/pygeomag/}{\texttt{pygeomag}}. The expected polarization angle is
\begin{equation}
    \phi_{p,\text{exp.}} = \sin^{-1}(|\,\hat{\mathbf{e}}_\text{geo}\times\hat{\mathbf{e}}_\phi\,|)~,
\end{equation}
where $\hat{\mathbf{e}}_\text{geo} =(\,\hat{\mathbf{v}}\times\hat{\mathbf{B}}\,)\,/\,|\,\hat{\mathbf{v}}\times\hat{\mathbf{B}}\,|$ and $\hat{\mathbf{e}}_\phi$ is the azimuthal basis vector of the spherical coordinate system.

Figure \ref{fig:polarization} compares the expected and reconstructed polarization angles, with uncertainties described in the next section.

\subsection{\label{sec:pol_unc}Polarization uncertainty estimates}

We estimate uncertainties for both the expected and measured polarization angles. For the measured angle, the full uncertainties result from three contributions, which we add in quadrature,
\begin{equation}
    \sigma_{\phi,\text{meas.}}^2 = \sigma_\text{reco}^2 + \sigma_\text{weak}^2 + \sigma_\text{ant}^2 ~,
\end{equation}
where $\sigma_\text{reco}$ quantifies the uncertainty in the reconstruction procedure due to the presence of thermal noise, $\sigma_\text{weak}$ quantifies the uncertainty from information imbalance due to an E-field component being buried within thermal noise, and $\sigma_\text{ant}$ quantifies the uncertainty in the antenna response due to differences between simplified ground planes and real terrain. 

The uncertainty in the expected polarization from pure geomagnetic emission is dominated by uncertainty in the reconstructed direction (see sections \ref{sec:rfrecon} and \ref{sec:screcon}), which determines the inferred shower axis and its orientation relative to the local magnetic field.

The effect from the charge-excess contribution on the polarization angle depends on both the relative strength and viewing angle off the shower axis, which we cannot constrain reliably. Thus, we include it using a profiled model statistic which we use only to compare against polarization reconstruction of events in background classes, as discussed in section \ref{sec:pol_bg}.

\subsubsection{\label{sec:unc_reco}Uncertainty on the reconstruction}

We estimate the uncertainty on our polarization reconstruction procedure by reconstructing synthetic events generated for each candidate CR event using the geomagnetic polarization angle expected from its reconstructed direction. These synthetic events combine a model impulsive E-field with 50 randomly selected thermal RF events from the SW-triggered dataset.

The synthetic events model the CR EAS emission as a 30 -- 80 MHz broadband E-field pulse forward-folded through each corresponding signal chain response (channel-specific) and antenna response (``top'' or ``bottom''; direction-specific). 

When adding to thermal RF events, the injection time for each synthetic pulse is randomly selected within 750 $\pm$ 50 ns to simulate timing jitter and evaluate the procedure's cross-correlation alignment. 

The synthetic events simulate the voltage SNRs measured in the observed events as follows. For each antenna, we take the target SNR to be the larger SNR of the Hpol and Vpol channels, and we build a set of realized SNRs by scaling the synthetic pulse over a range of trial scale factors before adding it to thermal noise waveforms. We choose the final scale factor to minimize the difference between the realized SNR and the target SNR, then scale both antenna polarization channels by it, simulating the polarization angle under the observed SNR. For SNR $< 4$, realized SNRs deviate from the target SNR on average by $\sim 0.3$, which is folded into the final uncertainty estimate.

As the synthetic events have simulated polarization angles equal to the geomagnetic polarization at the observed reconstructed direction, differences from the reconstructed polarization quantify the reconstruction statistical error, $\sigma_\text{reco}$. These uncertainties have a mean and standard deviation of $\langle\sigma_\text{reco}\rangle \simeq 3.12^\circ \pm 1.53^\circ$ across the 22 candidate CR events.

\subsubsection{\label{sec:unc_lowsnr}Uncertainty from a noise-dominated E-field component}

One of the E-field components used to reconstruct the polarization angle may be weak enough to be buried in thermal noise, particularly for nearly horizontal or vertical polarization. Varying the overall signal amplitude scales both components while preserving their ratio, and therefore the true polarization angle. However, while the weaker component remains noise-dominated, changes in the strength of the complementary component can bias the reconstructed polarization angle.

To estimate the uncertainty on the reconstructed polarization angle introduced by this signal imbalance, we fix the simulated polarization and build signals with a strong component over a range of SNR values. We reconstruct the polarization in each trial after adding thermal noise, and obtain distributions of the reconstruction residuals as a function of the realized SNR in the weak, noise-dominated component.

The uncertainty depends on both the weak component's SNR and the polarization angle,
\begin{equation}
    \sigma_\text{weak} = \sigma_\text{weak}(\text{weak SNR}, \phi_p)~.
\end{equation}
For each candidate CR event, we select the uncertainty curve corresponding to its reconstructed polarization angle and evaluate it at a representative SNR. We define this as the average SNR across all antennas for the E-field component with the lower-amplitude CSW. Among the 22 candidate CR events, this uncertainty has a mean and standard deviation of $\langle\sigma_\text{weak}\rangle \simeq 4.99^\circ \pm 2.92^\circ$.

\subsubsection{\label{sec:unc_ant}Uncertainty on the antenna response}

We presume that the dominant source of uncertainty in the antenna response comes from deviations between the true local ground geometry and the simplified ground planes used in the XFdtd response simulations described in section \ref{sec:antennas}. We consider only the response gain, and no additional phase uncertainty.

We parameterize the direction-dependent gain uncertainty with a scale factor, $\delta H$, which quantifies a characteristic magnitude of gain variation, and a dimensionless directional sensitivity factor, $\eta(\theta,\phi)$, which quantifies how strongly the ground model affects the antenna response at a given direction. The total uncertainty is then 
\begin{equation}
    \sigma_H(\theta,\phi) = \delta H\,\eta(\theta,\phi)~. \label{eqn:dH}
\end{equation}

We consider the three available XFdtd simulation configurations, ``ideal'', ``top'', and ``bottom'', as proxies for estimating these two factors. For $\delta H$, we compare the deviation between the ``top'' and ``bottom'' responses. For $\eta(\theta,\phi)$, we compare the deviation of the ``ideal'' response from the ``top'' and ``bottom'' responses. 

We compare responses in either case with a regularized fractional linear-gain difference,
\begin{equation}
    \Delta |H|_{A/B} = \frac{|H_A|-|H_B|}{\max\left[\frac{1}{2}\left(|H_A|+|H_B|\right), H_\text{floor}\right]}~,
    \label{eqn:regularized_gain_delta}
\end{equation}
where $A$ and $B$ can correspond to any of the response configurations, $H_\text{floor}$ is taken as 10\% of the combined RMS magnitudes and is calculated for each response component, \{$H_\theta^H$, $H_\phi^H$, $H_\theta^V$, $H_\phi^V$\}. This regularization prevents the antenna nulls from overestimating the gain differences. 

For $\delta H$, we apply eq. \eqref{eqn:regularized_gain_delta} to the ``top''/``bottom'' response comparison across all theta, phi, and frequencies. Then, we average the four components to estimate a single global value of the deviation magnitude, resulting in a value of $\sim 0.484$, which corresponds to $\sim 4$ dB deviations of the average gain over all directions. This overstates the effect on the reconstruction; the candidate CR events occupy regions with lower gain variation.

For the directional sensitivity factor, $\eta(\theta,\phi)$, we apply eq. \eqref{eqn:regularized_gain_delta} to the {``ideal''/``top''} and ``ideal''/``bottom'' response comparisons. For each response direction, we average $\eta(\theta,\phi)$ within a 5$^\circ$ patch, normalize it in the same manner as $\delta H$, and clip the values to (0.05, 2.5), which correspond to a total gain variation of (0.2 dB, 10 dB).

The largest uncertainties are near the antenna nulls and in the direction of the ground, where no candidate CR events reconstruct, but there is still an effect in their directions due to reflections. The values of $\eta(\theta,\phi)$ for candidate CR event directions range from 0.24 (1.0 dB total gain) to 1.06 (4.2 dB total gain).

We propagate the gain uncertainty of the antenna response to the reconstructed polarization with toy Monte Carlos similar to those in section \ref{sec:unc_reco}. To isolate the effect of the antenna response, the pulses are not added to thermal noise and SNR scaling is replaced with direct amplitude scaling corresponding to the trial polarization. Voltage waveforms are synthesized using the nominal antenna response, but the E-field is reconstructed using a randomly perturbed antenna response.

Then, we take the standard deviation of the reconstructed polarization errors as the contribution of the antenna response to the polarization angle uncertainty, $\sigma_\text{ant}$. Across the 22 candidate CR events, this gives a mean and standard deviation of $\langle\sigma_\text{ant}\rangle \simeq 3.05^\circ \pm 1.17^\circ$.

\subsubsection{\label{sec:unc_geo}Uncertainty on the expected geomagnetic polarization}

Uncertainties in pointing and direction of the shower axis dominate the uncertainty on the expected polarization angle. Note that the World Magnetic Model \cite{Chulliat2025WMM} reports uncertainties of $\sim 100$ nT for magnetic field strengths and 0.2$^\circ$ for inclination, which are negligible compared to the pointing uncertainties. 

To estimate the uncertainty from pointing, we compute the geomagnetic polarization angle at randomly sampled directions from a 2.5$^\circ$ wide Gaussian distribution centered at each $(\theta,\phi)$ on the sky to build distributions of the expected polarization angles. This gives a mean and standard deviation of the expected polarization angles at a given $(\theta,\phi)$. 

The resulting polarization angle is relatively stable for most of the sky, with a standard deviation less than 2$^\circ$. The standard deviation can rise to as much as 5$^\circ$ when the shower axis is nearly aligned with the magnetic field vector.

\subsubsection{\label{sec:ask}Estimate of a charge-excess contribution}

The radio emission from the EAS is dominated by geomagnetic emission but has a non-negligible charge-excess component that is radially polarized rather than having polarization determined by the local magnetic field. Proper consideration of the charge-excess contribution is usually done with a reconstructed E-field footprint, which cannot be constructed with the few densely packed antennas in the BEACON prototype. We instead consider the effect of possible charge-excess components that maximally deflect the polarization.

The relative strength of the charge-excess component is
\begin{equation}
    a = \sin(\alpha)\frac{|E_\text{ask}|}{|E_\text{geo}|}~, \label{eqn:a_ask}
\end{equation}
where $\alpha$ is the angle between the shower axis and magnetic field. Other experiments have simulated and measured $a$ for their detector sites \cite{auger_PhysRevD.89.052002, pol3_Schellart_2014, Chiche_2022, BELLETOILE201550}. A comprehensive comparison is given in ref. \cite{emission-Schr_der_2017}. Typical values for $a$ at the LOFAR site hover around 0.05 to 0.15. The magnetic field strength at the BEACON site is $48.2~\mu$T, close to LOFAR's $49~\mu$T, but BEACON's much higher altitude implies that observed EASs developed in lower atmospheric density. BEACON's geometry also admits a higher rate of inclined EASs, so combined with the higher altitude, we may expect a lower $a$.

We calculate the deflection of the pure geomagnetic polarization angle due to a charge-excess contribution with relative strength values $a = 0.05,~0.10,~0.15$ by adding geomagnetic and scaled charge-excess E-field vectors. Following eq. (8) from ref. \cite{auger_PhysRevD.89.052002}, the polarization angle of deflection, $\delta$, is
\begin{equation}
    \delta(\phi_\text{ask}) = \tan^{-1}\left(\frac{r\sin\phi_\text{ask}}{1 + r\cos\phi_\text{ask}}\right)~,
\end{equation}
where $r = a/\sin{\alpha}$. This function is maximized at the value
\begin{align}
    \cos(\phi_\text{ask, max}) &= -r \\
    \Rightarrow|\delta|_\mathrm{max} &= \sin^{-1}(r) = \sin^{-1}\left(\frac{a}{\sin\alpha}\right)~.
    \label{eqn:ask_def}
\end{align}

These maximum deflections are represented as allowed intervals in the expected polarization in figure \ref{fig:polarization}. They do not represent uncertainties --- the distributions of resultant polarization angles from adding randomly sampled radial-vector angles have standard deviations close to the maximum deflection value, but are heavily skewed, so we cannot add them in quadrature to the existing uncertainties.

\subsection{\label{sec:pol_bg}Polarization comparison to known background classes}

We compare the polarization reconstruction result for the 22 candidate CR events with that of various background classes. We choose 30 random events in each of four different background classes: BG-I: SW-triggered events with impulsive RF; BG-II: Scintillator-triggered events with impulsive RF and high RFI overlap scores; BG-III: Scintillator-triggered events with impulsive RF and low RFI overlap scores; and BG-IV: Scintillator-triggered events with RF-based analysis cuts, but no scintillator-based cuts (i.e. no direction agreement or saturation cut). The classes represent event samples that become progressively more CR-like. BG-I represents a truly random selection of background RFI. BG-II and BG-III represent clustered and non-clustered RF events coincident with scintillator signals. BG-IV represents a minimal RF-based CR analysis without scintillator reconstruction (except as an external trigger); it will contain CR events, but with a higher background rate.

We compare the reconstructed polarization angles of the candidate CR events and background classes using a reduced $\chi^2$,
\begin{equation}
    \chi_\text{red}^2 = \frac{1}{N}\sum_{i=1}^N \frac{(\phi_{p,\text{meas.}}^i - \phi_{p,\text{exp.}}^i)^2}{(\sigma^i_\text{meas.})^2 + (\sigma^i_\text{exp.})^2} ~.\label{eqn:chisq_pol1}
\end{equation}

This alone is sufficient to compare candidate CR events and background classes under a pure geomagnetic model ($a=0$). We can further include intervals allowed by charge-excess deflections by treating the deflection as a nuisance parameter over which we minimize residuals \cite{Rolke_2005, cousins2024phystat}. This is not a formal reduced-$\chi^2$ statistic, but can still be used to compare model agreement between the candidate CR events and the background classes under different possible strengths of a charge-excess contribution.

We define an additive deflection, $d$, within the range $[-\delta_a^i,~\delta_a^i]$, where $\delta_a^i$, calculated using eq. \eqref{eqn:ask_def}, is the maximum deflection caused by a charge-excess contribution with fixed strength parameter, $a$, for the $i^\text{th}$ point, which we then treat as a nuisance parameter. We extend eq. \eqref{eqn:chisq_pol1} with the added deflection, and profile the deflection over its allowed range by choosing the value of $d$ in $[-\delta_a^i,\delta_a^i]$ that minimizes the residual, resulting in a modified residual of 0 if the initial residual is smaller than $\delta_a^i$. This results in the profiled statistic,
\begin{equation}
    Q_a^2 = \frac{1}{N} \sum_{i=1}^N \frac{\left[\text{max}\{0,~|\phi_{p,\text{meas.}}^i - \phi_{p,\text{exp.}}^i|-\delta^i_a\}\right]^2}{(\sigma^i_\text{meas.})^2 + (\sigma^i_\text{exp.})^2} ~.
    \label{eqn:chisq_pol2}
\end{equation}
Larger values of $Q_a^2$ indicate the degree to which a set of events is unexplained by a charge-excess-caused deflection. This is a simplified method of including the charge-excess as a nuisance parameter, compared to profiling over a distribution, but it is suitable for comparisons of model agreement with different sets of events.

The resulting comparisons of the candidate CR event polarization reconstruction with those of the four background classes are shown in table \ref{tbl:pol_CRvsBG}. The absolute angular residual,
\begin{equation}
    r_a^i = \text{max}\{0,~|\phi_{p,\text{meas.}}^i - \phi_{p,\text{exp.}}^i|-\delta^i_a\}~,
\end{equation}
allows us to further compare the classes without weighting by large uncertainties.

\begin{table*}
    \centering
    \begin{tabular*}{\linewidth}{@{\extracolsep{\fill}}c|cccccc@{}}
        \toprule \toprule
        $a$ & CR $Q_a^2$ & CR $\langle r_a^i\rangle$ & BG-IV $Q_a^2$ & BG-IV $\langle r_a^i\rangle$ & BG Mean $Q_a^2$ & BG Mean $\langle r_a^i\rangle$\\
        \midrule
        0.00 & 3.04 & $8.8^\circ \pm 6.4^\circ$ & 2.68 & $14.7^\circ \pm 9.7^\circ$ & 9.46 & $13.6^\circ \pm 9.9^\circ$ \\
        0.05 & 1.56 & $5.6^\circ \pm 6.0^\circ$ & 1.65 & $10.8^\circ \pm 8.6^\circ$ & 7.19 & $10.3^\circ \pm 9.5^\circ$ \\
        0.10 & 0.72 & $3.2^\circ \pm 5.2^\circ$ & 0.90 & $7.5^\circ \pm 6.9^\circ$ & 5.44 & $7.8^\circ \pm 8.8^\circ$ \\
        0.15 & 0.32 & $1.8^\circ \pm 4.1^\circ$ & 0.42 & $4.5^\circ \pm 5.3^\circ$ & 4.02 & $5.9^\circ \pm 7.8^\circ$ \\
        \bottomrule
    \end{tabular*}
    \caption{The charge-excess-profiled residual statistics, $Q_a^2$, and average absolute angular residuals, $\langle r_a^i\rangle$, comparing the polarization reconstruction of the 22 candidate CR events to the background-sampled classes defined in the text. The row with no charge-excess contribution, $a=0.00$, is equivalent to the pure geomagnetic residual statistic without profiling. The table separately lists the best-performing class, BG-IV, and the BG-I -- BG-III classes (``BG'' here) by their mean values.}
    \label{tbl:pol_CRvsBG}
\end{table*}

The polarization reconstruction provides a consistency test for expected CR radio emission. The polarization reconstruction for the candidate CR events has substantially smaller profiled residual statistics than the mean of the generic RF-background classes, BG-I to BG-III, for every tested charge-excess strength parameter, $a$. As an example, in the pure geomagnetic case, the candidate CR events give $Q_{a=0}^2=3.04$, while the BG-I to BG-III mean is $Q_{a=0}^2=9.46$.

The background class, BG-IV, is a lower-purity sample of CR events, with identical RF-selection criteria as the 22 candidate CR events. Therefore, BG-IV is unsurprisingly the best-performing background class and is in fact comparable to the candidate CR events in the profiled statistic. For the pure geomagnetic comparison, BG-IV gives $Q_{a=0}^2=2.68$, slightly smaller than the candidate set value of $Q_{a=0}^2=3.04$. Despite this, the comparable weighted residual statistic is explained by BG-IV containing overall larger deflections and estimated uncertainties compared to the candidate CR events.

Overall, the candidate CR events have smaller angular residuals, smaller reconstruction uncertainties, and stronger separation from the generic RF-background classes. BG-IV remains comparable in the weighted statistic, but its larger angular spread and uncertainties indicate that it is less tightly matched to the expected polarization model. These results support the interpretation that the candidate CR events have reconstructed polarization angles compatible with the expected polarization of CR radio emission.



\section{\label{sec:conc}Conclusions}

A data-driven selection of scintillator-coincidence-triggered events containing impulsive RF signals observed by the 2023 BEACON prototype identifies 22 candidate CR events on an estimated background of 0.11 events. An independent cross-check against recorded airplane tracks finds two candidate CR events within a conservative $5^\circ$ pointing bound with a 0.9$\sigma$ significance, but none within $2.5^\circ$. We exclude no events on this basis.

We further characterize the candidate CR events through local E-field and polarization reconstruction. We find that their reconstructed polarization angles agree with a geomagnetic emission model with a charge-excess contribution represented as allowed intervals of polarization angle deflection, and that these agree better with expected CR polarization than generic RF-background classes do. However, the polarization comparison alone is not sufficient to distinguish the candidate CR events from the most CR-like RF-selected background class, particularly because of reconstruction uncertainties and the unconstrained charge-excess contribution.

Overall, these results demonstrate that a scintillator array can provide an effective cross-validation channel for radio-detectable CR events in the 2023 BEACON prototype. The results also suggest that the RF array can detect CR EAS radio emission; however, pervasive RFI in the BEACON environment makes such emission difficult to distinguish during analysis.

We suggest several directions for improving CR EAS discrimination from RFI backgrounds with future elevated radio arrays. A sparse array surrounding the phased station could provide a reliable estimate of the shower LDF, improving the characterization of CR radio polarization by constraining the charge-excess contribution, providing an additional means of rejecting RFI, and enabling reconstruction of the primary energy and depth of shower maximum, $X_\text{max}$ \cite{emission-Schr_der_2017}. Additional antennas would also help reduce sidelobe ambiguities in direction reconstruction, which caused many RFI events to reconstruct into secondary clusters above the horizon. Another improvement would be better characterization of the antenna response uncertainty, perhaps using a suitably equipped drone~\cite{augerOctocopter}. Additional simulations would then only be required to quantify uncertainties from seasonal changes in the dielectric properties of the ground. Better knowledge of the antenna response could also allow us to incorporate additional information from simulations, such as signal polarity.

Finally, the reconstructed E-field amplitudes span an order of magnitude among the candidate CR events; the mean value is $\sim 300\,\mu$V/m, but some events reach $\sim 3000\,\mu$V/m, which may motivate increasing the digitizer dynamic range to avoid RF-signal saturation of the strongest CR EAS emission.

In August 2025, the BEACON prototype was decommissioned. A future elevated array concept, the Hybrid Elevated Radio Observatory for Neutrinos (HERON), is planned to combine BEACON-like elevated phased arrays with GRAND-like sparse autonomous antenna arrays deployed along a mountain range overlooking a valley in Argentina's San Juan province \cite{heron}. Such a hybrid design could preserve the low-threshold advantages of phased-array triggering while adding the footprint information needed for stronger CR reconstruction.


\acknowledgments
This work is supported by NSF Awards $\#$ 2033500, 1752922, 1607555, and DGE-1746045 as well as the Sloan Foundation, the RSCA, and NASA (support through JPL and Caltech as well as Award $\#$ 80NSSC18K0231), and by the Gordon and Betty Moore Foundation, grant DOI 10.37807/GBMF11570.

This work has received financial support from Ministerio de Ciencia, Innovaci\'on y Universidades/Agencia Estatal de Investigaci\'on, MICIU/AEI /10.13039/501100011033, Spain
(PID2022-140510NB-I00, PCI2023-145952-2, PID2025-170557NB-I00 and Mar\'\i a de Maeztu grant CEX2023-001318-M); Xunta de Galicia, Spain (CIGUS Network of Research Centers, and
Consolidaci\'on ED431C-2025/11); and the European Union (Galicia Feder 2021-2027 Program and Horizon Europe Research and Innovation Program under Grant Agreement No. 101131928 - ACME).

We thank the NSF-funded White Mountain Research Center and Steven DeVanzo for their support. Computing resources were provided by the University of Chicago Research Computing Center, the Institute for Gravitation and the Cosmos and the Institute for Computational and Data Sciences at Penn State. We further acknowledge support from KIT for donating scintillator panels.

\bibliographystyle{JHEP}
\bibliography{sources}

\end{document}